\documentstyle[11pt]{article}

\begin{document}
\title{ $SU(3)$ Yang-Mills Hamiltonian in the flux-tube gauge:\\ Strong coupling expansion and glueball dynamics}
\author{Hans-Peter Pavel \\[1cm]
Bogoliubov Laboratory of Theoretical Physics,
\\
Joint Institute for Nuclear Research, Dubna, Russia\\
and\\
        Institut f\"ur Physik und Institut f\"ur Mathematik,  Humboldt-Universit\"at zu Berlin, IRIS Haus,\\ 
        Zum Grossen Windkanal 6, D-12489 Berlin, Germany\footnote{email: hans-peter.pavel@physik.hu-berlin.de, pavel@theor.jinr.ru}}
\date{December 14th, 2021}
\maketitle

\begin{abstract}
It is shown that the formulation of the $SU(3)$ Yang-Mills quantum Hamiltonian in the "flux-tube gauge"  
 $$A_{a1}=0\quad {\rm for\ all}\quad  a=1,2,4,5,6,7 \quad\quad {\rm and}  \quad\quad A_{a2}=0\quad {\rm for\ all}\quad  a=5,7~, $$
allows for a systematic and practical strong coupling expansion of the Hamiltonian in $\lambda\equiv g^{-2/3}$, 
equivalent to an expansion in the number of spatial derivatives.
Introducing an infinite spatial lattice with box length $a$, the "free part" is the sum of Hamiltonians of Yang-Mills quantum mechanics
of constant fields for each box, and the "interaction terms" contain higher and higher number
of spatial derivatives connecting different boxes. 
The Faddeev-Popov operator, its determinant and inverse, are rather simple, but 
show a highly non-trivial periodic structure of six Gribov-horizons separating six Weyl-chambers. 
The energy eigensystem of the gauge reduced Hamiltonian of $SU(3)$ Yang-Mills mechanics of spatially constant
fields can be calculated in principle with arbitrary high precision using the orthonormal basis of all solutions of the corresponding
harmonic oscillator problem, which turn out to be made of orthogonal polynomials of the 45 components of eight irreducible
symmetric spatial tensors. First results for the low-energy glueball spectrum are obtained which substantially improve those
by Weisz and Ziemann using the constrained approach.
Thus, the  gauge reduced approach using the flux-tube gauge proposed here, is expected to enable one to obtain 
valuable non-perturbative information about low-energy glueball dynamics,
using perturbation theory in $\lambda$.
\end{abstract}

\section{Introduction}

The Hamiltonian approach \cite{Christ and Lee}, with the possibility to use the powerful variational method,
has turned out to be very suitable for non-perturbative investigations of Yang-Mills theory. 

In this work, the Yang-Mills theory for \(SU(3)\) gauge fields \( V^a_{\mu}(x) \) is considered, defined by the action
\begin{equation}
{\cal S} [V] : = - \frac{1}{4}\ \int d^4x\ F^a_{\mu\nu} F^{a\mu \nu}~,\quad
F^a_{\mu\nu} : = \partial_\mu V_\nu^a  -  \partial_\nu V_\mu^a
+ g\, f^{abc} V_\mu^b V_\nu^c~,\nonumber
\end{equation}
invariant under Poincar\'{e} and scale transformations,
and under local $SU(3)$ gauge transformations
$U[\omega(x)]\equiv \exp(i\omega_a\lambda_a/2)$
\begin{equation}
V_{a\mu}^{\omega}(x) \lambda_a/2   =
U[\omega(x)] \left(V_{a\mu}(x) \lambda_a/2 +
{i\over g}\partial_\mu \right) U^{-1}[\omega(x)]~.
\nonumber
\label{4-gauge-trafo}
\end{equation}

The corresponding quantum theory is then obtained
by exploiting the time dependence of the gauge transformations to put
\begin{eqnarray}
V_{a0}(x) = 0~,\quad a=1,...,8~,\nonumber \quad\quad (\rm{Weyl}\  \rm{Gauge})\nonumber
\end{eqnarray}
and imposing canonical commutation relations on the spatial components of the gauge fields and (the negative of) the
chromoelectric fields, using the Schr\"odinger representation
\begin{eqnarray}
 [\Pi_{ai}({\mathbf{x}}),V_{bj}({\mathbf{y}})]=i\delta_{ab}\delta_{ij}\delta({\mathbf{x}}-{\mathbf{y}})\quad \longrightarrow
\quad \Pi_{ai}({\mathbf{x}})=-E_{ai}({\mathbf{x}})=-i\delta/\delta V_{ai}({\mathbf{x}})~.\nonumber
\end{eqnarray}
The physical states $\Psi$ have to satisfy the system of equations (see e.g. \cite{Christ and Lee})
\begin{eqnarray}
&&(H-E)\Psi = 0\quad ({\rm Schr\ddot{o}dinger}\ {\rm equation})~,\nonumber\\
&&G_a({\mathbf{x}})\Psi = 0\quad ({\rm Gauss}\ {\rm law}\ {\rm constraints})~,\label{System}
\end{eqnarray}
with the Hamiltonian
\begin{equation}
H = \int d^3{\mathbf{x}} {1\over 2}
\sum_{a,i} \left[ \Pi_{ai}^2({\mathbf{x}})+B_{ai}^2(V({\mathbf{x}}))\right]~,
\nonumber
\end{equation}
in terms of the chromoelectric fields $-\Pi_{ai}$ and the chromomagnetic $  B_{ai}(V) =\epsilon_{ijk}\left(\partial_j V_{ak}+{1\over 2}g
f_{abc}V_{bj} V_{ck}\right)$, and the Gauss law operators
\begin{equation}
G_a ({\mathbf{x}}) = -i\left(\delta_{ac}\partial_i +g\, f_{abc}V_{bi}({\mathbf{x}})\right)
            \Pi_{ci}({\mathbf{x}})~,\quad a=1,...,8~,\nonumber
\end{equation}
which are the generators of the residual time independent gauge transformations,
commute with the Hamiltonian and satisfy angular momentum commutation relations
\begin{eqnarray}
[G_a({\mathbf{x}}),H]=0~,\quad\quad [G_a({\mathbf{x}}),G_b({\mathbf{y}})]=ig\delta({\mathbf{x}}-{\mathbf{y}})
 f_{abc}G_c({\mathbf{x}})~,\nonumber
\end{eqnarray}
The matrix elements are Cartesian
\begin{equation}
\langle\Psi_1|{\cal O} |\Psi_2\rangle=
\int \prod_{ai} dV_{ai} \ \Psi_1^* {\cal O}\Psi_2~.\nonumber
\end{equation}

In order to calculate the eigenstates and their energies, it is useful to implement
the non-Abelian Gauss law constraints into the Schr\"odinger equation
by further fixing the gauge using the remaining time-independent gauge transformations.

One possibility, well suited for the high energy sector of the theory, is to
impose the Coulomb gauge $\chi_a(A)=\partial_i A_{ai}=0$ describing the dynamics in terms
of physical colored transverse gluons. 
For example, integrating out all higher modes in a small box of size a,  L\"uscher \cite{Luescher} 
 obtained a weak coupling expansion for the energies
of the constant Yang-Mills fields, $E = \frac{1}{a}\sum_{k=0}^\infty \epsilon_k \bar{\lambda}^k~,
\  \bar{\lambda}\equiv[\bar{g}(\Lambda_{MS}a)]^{2/3}$, with the standard running coupling
constant in the MS scheme.
Already some time ago, Weisz and Ziemann\cite{Weisz and Ziemann}, when applying L\"uscher's results to the case of $SU(3)$,
 obtained rather accurate values for the eigensystem of $SU(3)$ Yang-Mills quantum mechanics of spatially constant
fields working directly in the unreduced V-space discussed above.

For the low-energy sector of $SU(2)$ Yang-Mills theory, the symmetric gauge \cite{KP1} has been proven to exist \cite{KMPR}
and was shown in \cite{pavel2010} to be very well suited to describe non-perturbative glueball dynamics 
by demonstrating how the symmetric gauge allows for a gauge invariant formulation of $SU(2)$ Yang-Mills theory on a
three dimensional spatial lattice by  replacing integrals by sums and spatial derivatives by differences. 
Using the symmetric gauge and constructing the corresponding physical quantum Hamiltonian of $SU(2)$ Yang-Mills theory
according to the general scheme given by Christ and Lee \cite{Christ and Lee},
it was proven in  \cite{pavel2010} that a strong coupling expansion of the $SU(2)$ Yang-Mills quantum Hamiltonian
in  $\lambda=g^{-2/3}$ can be carried out equivalent to an expansion in the number of spatial derivatives.
Introducing an infinite lattice with box length $a$,
a systematic strong coupling expansion of the Hamiltonian in $\lambda$ can be obtained,
with the "free part" being the sum of Hamiltonians of Yang-Mills quantum mechanics
of constant fields for each box, and "interaction terms" of higher and higher number
of spatial derivatives connecting different boxes.
Using the very accurate results of Yang-Mills quantum mechanics of constant fields in a box,
obtained with the variational method in earlier work \cite{pavel2007},
the energy spectrum and the dynamics of weakly interacting glueballs can be calculated systematically and with high accuracy,
using perturbation theory in $\lambda$.
The coarse graining approach can also straight forwardly be generalised to the inclusion of fermions
using the more recent results on the $SU(2)$ Dirac-Yang-Mills quantum mechanics of spatially constant
quark- and gluon-fields \cite{pavel2011}.
It offers a useful alternative to lattice calculations based on the Wilson-loop \cite{Creutz},
including the corresponding analytic strong coupling expansions by Kogut, Sinclair, and Susskind \cite{Kogut}
and M\"unster \cite{Muenster} and the calculation of the glueball spectrum (e.g.\cite{Morningstar},\cite{Chen}).

The generalisation of the symmetric gauge from $SU(2)$ Yang-Mills theory to $SU(3)$, proposed in \cite{pavel2012}-\cite{pavel2014},
indeed leads to a unconstrained dynamical description in terms of 16 colorless variables with definite spin quantum numbers,
spin-0, spin-1, spin-2, and spin-3, and in principle allows for an expansion of the Hamiltonian in spatial derivatives, 
but due to an algebraically highly intricate FP-operator, is not practically managable at the moment.

In this work a new gauge, the flux-tube gauge, is proposed, which exists for low-energy $SU(3)$ Yang-Mills theory,
and allows for a derivative expansion with a rather simple, but non-trivial Faddeev-Popov operator, showing a 6-fold
singularity structure with six Gribov horizons separating six Weyl-chambers. Here, the rather accurate recent results for the
low-energy eigensystem of SU(3) Yang-Mills Quantum Mechanics \cite{pavel2021} are used,
substantially improving the results by Weisz and Ziemann \cite{Weisz and Ziemann}.

The article is organised as follows: In Sec.2,  the flux-tube gauge is defined and the corresponding unconstrained quantum Hamiltonian
presented. In Sec.3, the expansion of the Hamiltonian up to second order in the number of spatial derivatives is carried out.
Sec.4 revises the coarse grainig method and the strong coupling expansion in  $\lambda=g^{-2/3}$. 
In Sec.5, the lowest order, $SU(3)$ Yang-Mills quantum mechanics of spatially constant fields in a box, and
the calculation of its energy spectrum are discussed. Sec.6 outviews how the perturbation theory in $\lambda$
can be carried out, and Sec.7 states the conclusions. In the Appendix the proof of the existence of the flux-tube gauge for strong coupling
is given.

\section{Physical $SU(3)$ Quantum Hamiltonian in the flux-tube gauge}

\subsection{Definition of the flux-tube gauge}

I shall here choose the flux-tube gauge
\begin{equation}
A_{a1}=0\quad\quad \forall a=1,2,4,5,6,7 \quad\quad \wedge  \quad\quad A_{a2}=0\quad\quad \forall a=5,7~.
\label{gauge cond}
\end{equation}
In the standard notation of \cite{{Christ and Lee}} this gauge condition can be written in the form
\begin{equation}
\chi_a(A)=\left(\Gamma_i\right)_{ab}\, A_{bi}=0~,\quad\quad a=1,...,8~.
\label{GF1}
\end{equation}
The explicit form of the $8\times 8$-matrices $\Gamma_{i}$ which satisfy $ \Gamma_{i}^T\Gamma_{i}=1$ is 
\begin{equation}
\Gamma_{i}\equiv
\delta_{i1}\left(
\begin{array}{c c c c c c c c}
1&0&0&0&0&0&0&0 \\
0&1&0&0&0&0&0&0 \\
0&0&0&0&0&0&0&0 \\
0&0&0&1&0&0&0&0 \\
0&0&0&0&1&0&0&0 \\
0&0&0&0&0&1&0&0 \\
0&0&0&0&0&0&1&0 \\
0&0&0&0&0&0&0&0
\end{array}
\right)+
\delta_{i2}\left(
\begin{array}{c c c c c c c c}
0&0&0&0&0&0&0&0 \\
0&0&0&0&0&0&0&0 \\
0&0&0&0&1&0&0&0 \\
0&0&0&0&0&0&0&0 \\
0&0&0&0&0&0&0&0 \\
0&0&0&0&0&0&0&0 \\
0&0&0&0&0&0&0&0 \\
0&0&0&0&0&0&1&0
\end{array}
\right)~.
\label{GF2}
\end{equation}
The existence of the flux-tube gauge (\ref{gauge cond}) for SU(3) Yang-Mills theories in $D+1$ dimensions with $D\ge 2$ spatial 
dimensions, can be proven by construction for the case of spatially constant fields\footnote{This is the generalisation of the
$SU(2)$ gauge fixation introduced by Green and Gutperle \cite{Green and Gutperle}, first rotating the spatial 1-component
into the color-3 direction, $A^{\rm SU2}_{a1}=0$ for all $a=1,2$, and then use remaining color-rotations orthogonal to the $a=3$ direction, 
generated by $\sigma_3$, which leave the $a=3$ components of all $A^{\rm SU2}_{ai}$ for all $i=1,2,3$ 
unchanged, in order to put in the spatial 2-component $A^{\rm SU2}_{a2}=0$ for $a=1$.}: 
One first diagonalises the spatial 1-component of the gauge
field in the fundamental representation, $A_{a1}=0$ for all $a=1,2,4,5,6,7$. Then, one uses the remaining
gauge-freedom, generated by $\lambda_3$ and $\lambda_8$, which leave the $a=3,8$ components of all $A_{ai}$ for all $i=1,2,3$ 
unchanged,  to put the spatial 2-components $A_{a2}=0$ for $a=5,7$. 
Such a construction is unique for field- configurations, for which the corresponding homogeneous Faddeev-Popov determinant, 
to be discussed in Sect. 2.3., is non-vanishing.
A more detailed discussion of the existence of the flux-tube gauge and 
the extension of the proof  to the case of spatially slightly varying fields in the sense of a strong-coupling expansion,
is given in App. A  "On the existence of the flux-tube gauge".

The flux-tube gauge corresponds to the
point transformation to the new set of adapted coordinates,
the 8 $ q_j\ \ (j=1,...,8)$  and the 16 elements $A_{ai}$ 
\begin{equation}
\label{coordtrafo}
V_{ai} \left(q, A \right) =
O_{ab}\left(q\right) A_{bi}
- {1\over 2g}\, f_{abc} \left( O\left(q\right)
\partial_i O^T\left(q\right)\right)_{bc}\,,\nonumber
\end{equation}
where \( O(q) \) is an orthogonal $8\times 8$ matrix adjoint to $U(q)$ parametrized by the 8 $q_i$ and the physical fields
\begin{equation}
A=\,
 \left(
\begin{array}{c c c}
 0 & A_{12} & A_{13} \\ 
 0 & A_{22} & A_{23}\\ 
 A_{31} & A_{32} & A_{33} \\ 
 0 & A_{42} & A_{43} \\ 
 0 & 0 & A_{53}\\ 
 0 & A_{62} & A_{63} \\ 
 0 & 0 & A_{73} \\ 
 A_{81} & A_{82} & A_{83}    
\end{array}
\right)
\equiv  
\left( X\   Y\   Z  \right)~.  
\end{equation}

\subsection{Physical $SU(3)$ Quantum Hamiltonian}

After the above coordinate transformation (\ref{coordtrafo}), the non-Abelian
Gauss-law constraints become the Abelian conditions
\begin{eqnarray}
 G_a\Psi=0\quad \Leftrightarrow\quad\frac{\delta }{\delta q_i}\Psi=0\quad
 (\rm{Abelianisation}),
 \nonumber
\end{eqnarray}
that the physical states should depend only on the physical variables $A_{ik}$, and the system
(\ref{System})
reduces to the unconstrained Schr\"odinger equation
\begin{equation}
H(A,P)\Psi(A)=E\Psi(A)~.
\end{equation}

The correctly ordered physical quantum Hamiltonian \cite{Christ and Lee} in the flux-tube gauge
in terms of the physical variables $A_{ik}({\mathbf{x}})$  and the corresponding canonically
conjugate momenta
$P_{ik}({\mathbf{x}})\equiv -i\delta /\delta A_{ik}({\mathbf{x}})$  reads
\begin{eqnarray}
H(A,P)\!\!\!\!&=&\!\!\!\! {1\over 2}{\cal J}^{-1}\!\!\int d^3{\mathbf{x}}\int d^3{\mathbf{y}}
\ P_{ai}({\mathbf{x}})\ {\cal J}\ {\cal K}_{ai|bk}({\mathbf{x}},{\mathbf{y}})\ P_{bk}({\mathbf{y}})
     +{1\over 2}\int d^3{\mathbf{x}}\left(B_{ai}(A)\right)^2~,
\end{eqnarray}
with the kernel
\begin{equation}
\label{kernel}
{\cal K}_{ai|bk}({\mathbf{x}},{\mathbf{y}})\equiv
            \delta_{ab}\delta_{ik}\delta({\mathbf{x}}-{\mathbf{y}})
            -\langle {\mathbf{x}}\ a|D_i(A)\
      ^{\ast}\! D(A)^{-1}\,  ^{\ast}\! D^{\dagger}(A)^{ -1} \ D_k(A)|{\mathbf{y}}\ b\rangle~,
\end{equation}
the Jacobian
\begin{equation}
\label{J}
{\cal J}\equiv\det |^{\ast}\! D(A)|~.
\end{equation}
Here $D_i(A)$ is the antisymmetric operator whose matrix elements are the covariant derivatives
\begin{eqnarray}
\langle {\mathbf{x}}\ a| D_i(A)|{\mathbf{y}}\ b\rangle
=  D_i (A)_{ab}^{({\mathbf{x}})}\delta({\mathbf{x}}-{\mathbf{y}})
\equiv
\left(\delta_{ab}\partial_i^{({\mathbf{x}})}-g\, f_{abc}A_{ci}({\mathbf{x}})\right)\delta({\mathbf{x}}-{\mathbf{y}})~,
\nonumber
\end{eqnarray}
and $^{\ast}\! D(A)$ the Faddeev-Popov (FP) operator whose matrix elements are defined as 
\begin{equation}
\label{FP}
\langle {\mathbf{x}}\ a|^{\ast}\!  D(A)|{\mathbf{y}}\ b\rangle =
^{\ast}\!\!\! D(A)_{ab}^{({\mathbf{x}})}\delta({\mathbf{x}}-{\mathbf{y}})
\equiv \left(\Gamma_i\right)_{ac}\, D_i(A)^{({\mathbf{x}})}_{cb}\delta({\mathbf{x}}-{\mathbf{y}})
   =\left(\left(\Gamma_{i}\right)_{ab}\partial_i^{({\mathbf{x}})}-g\gamma_{ab}({\mathbf{x}})\right)\delta({\mathbf{x}}-{\mathbf{y}})~,
\end{equation}
with the homogeneous part of the FP operator
\begin{eqnarray} 
\gamma_{ab}({\mathbf{x}})\equiv \left(\Gamma_i\right)_{ad} \, f_{dbc}A_{ci}({\mathbf{x}})~.
\label{homogFP}
\end{eqnarray}
Its explicit expression is
\begin{equation}
\gamma =
 \left(
\begin{array}{c c c c c c c c}
0  & -X_3 & 0 & 0 & 0 & 0 & 0 & 0 \\ 
X_3 & 0 & 0 & 0 & 0 & 0 & 0 & 0 \\ 
-Y_6/2 & 0 & -Y_4/2 &-Y_+  & 0 & Y_1/2 & -Y_2/2 &  - \sqrt{3}\, Y_4/2 \\ 
0 & 0 & 0 & 0 &X_+ & 0 & 0 & 0 \\ 
0 & 0 & 0 &-X_+  & 0 & 0 & 0 & 0 \\ 
0 & 0 & 0 & 0 & 0 & 0 &-X_-  & 0 \\ 
0 & 0 & 0 & 0 & 0 & X_-  & 0 & 0 \\ 
-Y_4/2 & 0 & Y_6/2 & Y_1/2 & Y_2/2 & Y_- & 0 & - \sqrt{3}\, Y_6/2  
\label{gamma}
\end{array}
\right)~,  
\end{equation}
using the abbreviations $X_\pm:=-(X_3\pm \sqrt{3}X_8)/2$ and
 $Y_\pm:=-(Y_3\pm \sqrt{3}Y_8)/2$. 

\noindent
The matrix element of a physical operator $O$ is given by
\begin{equation}
\label{ME}
\langle \Psi_1| O|\Psi_2\rangle\
\propto
\int \prod_{\mathbf{x}}\Big[dA({\mathbf{x}})\Big]
{\cal J}\Psi_1^*[A] O\Psi_2[A]~.
\end{equation}

\subsection{Existence of an expansion in the number of spatial derivatives}

The Green functions of the FP operator  (\ref{FP}),
and hence the non-local terms of the physical Hamiltonian, can
-in contrast for example to the Coulomb gauge- be expanded in the number of spatial derivatives.
\begin{eqnarray}
\langle {\mathbf{x}}\ a| ^{\ast}\! D(A)^{-1}|{\mathbf{y}}\ b\rangle &=&
-{1 \over g}\left(\gamma^{-1}({\mathbf{x}})\right)_{ac}
\Bigg[ \delta_{c b}\delta({\mathbf{x}}-{\mathbf{y}})
+{1 \over g}\left( \Gamma_i\right)_{cd}
     \partial_i^{({\mathbf{x}})}
     \left[\left(\gamma^{-1}({\mathbf{x}})\right)_{db}\delta({\mathbf{x}}-{\mathbf{y}})\right]
\nonumber\\
&&\quad\quad\quad\quad\quad 
+{1 \over g^2}\left( \Gamma_i\right)_{cd}
     \partial_i^{({\mathbf{x}})}\!\!\left[\left(\gamma^{-1}({\mathbf{x}})\right)_{de}\left( \Gamma_j\right)_{en}
     \partial_j^{({\mathbf{x}})}\!\!
     \left[\left(\gamma^{-1}({\mathbf{x}})\right)_{nb}\delta({\mathbf{x}}-{\mathbf{y}})\right]\right]
+...\Bigg]~.
\nonumber
\end{eqnarray}
The inverse $\gamma^{-1}$ of the homogeneous part of the Faddeev-Popov operator (\ref{gamma}) exists in the regions of non-vanishing
 determinant  
\begin{equation}
{\rm det}\,\gamma=X^2_3\left(X^2_3-3\,X^{2}_8\right)^2 Y_4 Y_6
\label{det-gamma-hom}
\end{equation}
and the non-vanishing matrix elements of the  inverse $\gamma^{-1}$ are rather simple,
\begin{eqnarray}
&&
\left(\gamma^{-1}\right)_{12}=-\left(\gamma^{-1}\right)_{21}=X_3^{-1}~,  
\nonumber\\
&&
\left(\gamma^{-1}\right)_{32}={1\over 2}X_3^{-1} \left(Y_4/Y_6-Y_6/Y_4\right)~,
\quad 
\left(\gamma^{-1}\right)_{82}=-{1\over 2\sqrt{3}}X_3^{-1} \left(Y_4/Y_6+Y_6/Y_4\right)~,
\nonumber\\
&&---------------------------------------
\nonumber\\
&&
\left( \gamma^{-1}\right)_{45}=-\left(\gamma^{-1}\right)_{54}=-X_+^{-1}~,
\quad\quad  
\left(\gamma^{-1}\right)_{34}=-\sqrt{3}\left(\gamma^{-1}\right)_{84}=-{1\over 2} X_+^{-1}(Y_2/Y_6)~,
\nonumber\\
&&
\left(\gamma^{-1}\right)_{35}=-{1\over 2} X_+^{-1}\left(-Y_1/Y_6-2\,Y_+/Y_4 \right)~,
\quad\quad  
\left(\gamma^{-1}\right)_{85}=-{1\over 2\sqrt{3}}X_+^{-1}\left(Y_1/Y_6-2\,Y_+/Y_4 \right)~,
\nonumber\\
&&---------------------------------------
\nonumber\\
&&
\left( \gamma^{-1}\right)_{67}=-\left(\gamma^{-1}\right)_{76}=X_-^{-1}~,
\quad\quad  
\left( \gamma^{-1}\right)_{36}=\sqrt{3}\left(\gamma^{-1}\right)_{86}={1\over 2}X_-^{-1}(Y_2/Y_4)~,
\nonumber\\
&&
 \left( \gamma^{-1}\right)_{37}={1\over 2}X_-^{-1}\left(Y_1/Y_4-2\,Y_-/Y_6 \right)~,
\quad\quad  
\left(\gamma^{-1}\right)_{87}={1\over 2 \sqrt{3}}X_-^{-1}\left(Y_1/Y_4+2\,Y_-/Y_6 \right)~,
\nonumber\\
&&---------------------------------------
\nonumber\\
&&
\left(\gamma^{-1}\right)_{33}=\sqrt{3}\left(\gamma^{-1}\right)_{83}=Y_4^{-1}~,
\quad\quad  
\left(\gamma^{-1}\right)_{38}=-\sqrt{3}\left(\gamma^{-1}\right)_{88}=Y_6^{-1}~,
\label{gammainv}
\end{eqnarray}
grouped into those proportional $X_3^{-1}$, $X_+^{-1}$, and $X_-^{-1}$, respectively
and those independent of $X$. Such a "Weyl-decomposition" will lead to a considerable simplification
of the non-local part of the kernel  (\ref{kernel}).
From the definition of the gauge-fixing (\ref{GF2}) we have
\begin{eqnarray}
\left( \Gamma_i\right)_{ac}
     \partial_i^{({\mathbf{x}})}
     \left[\gamma^{-1}_{cb}({\mathbf{x}})\delta({\mathbf{x}}-{\mathbf{y}})\right]=
\left( \Gamma_1\right)_{ac}
     \partial_1^{({\mathbf{x}})}
     \left[\gamma^{-1}_{cb}({\mathbf{x}})\delta({\mathbf{x}}-{\mathbf{y}})\right]
+\left( \Gamma_2\right)_{ac}
     \partial_2^{({\mathbf{x}})}
     \left[\gamma^{-1}_{cb}({\mathbf{x}})\delta({\mathbf{x}}-{\mathbf{y}})\right]~.
\end{eqnarray}
Using the values (\ref{gammainv}), we find
\begin{eqnarray}
\left( \Gamma_1\right)_{ac}
     \partial_1^{({\mathbf{x}})}
     \left[\gamma^{-1}_{cb}({\mathbf{x}})\delta({\mathbf{x}}-{\mathbf{y}})\right]
&=&{\cal Q}^{(0)}_{ab}
     \partial_1^{({\mathbf{x}})}
     \left[{1\over X_3({\mathbf{x}})}\delta({\mathbf{x}}-{\mathbf{y}})\right]
+{\cal Q}^{(+)}_{ab}
     \partial_1^{({\mathbf{x}})}
     \left[{1\over X_+({\mathbf{x}})}\delta({\mathbf{x}}-{\mathbf{y}})\right]
\nonumber\\
&&\quad\quad
+{\cal Q}^{(-)}_{ab} 
     \partial_1^{({\mathbf{x}})}
     \left[{1\over X_-({\mathbf{x}})}\delta({\mathbf{x}}-{\mathbf{y}})\right]~,
\end{eqnarray}
with the $8\times 8$  matrices
\begin{eqnarray}
{\cal Q}^{(0)}_{ab}=\delta_{a\,[ 1}\delta_{b\, 2]}   \quad\quad\quad  {\cal Q}^{(+)}_{ab}=-\delta_{a\,[ 4}\delta_{b\, 5]}   \quad\quad\quad
{\cal Q}^{(-)}_{ab}=\delta_{a\,[ 6}\delta_{b\, 7]} ~,  
\end{eqnarray}
and
\begin{eqnarray}
\left( \Gamma_2\right)_{ac}
     \partial_2^{({\mathbf{x}})}
     \left[\gamma^{-1}_{cb}({\mathbf{x}})\delta({\mathbf{x}}-{\mathbf{y}})\right]
&=& {\cal Q}^{(3)}_{ab}
     \partial_2^{({\mathbf{x}})}
     \left[{1\over X_+({\mathbf{x}})}\delta({\mathbf{x}}-{\mathbf{y}})\right]
+{\cal Q}^{(8)}_{ab}
     \partial_2^{({\mathbf{x}})}
     \left[{1\over X_-({\mathbf{x}})}\delta({\mathbf{x}}-{\mathbf{y}})\right]~,
\end{eqnarray}
with the $8\times 8$  matrices
\begin{eqnarray}
{\cal Q}^{(3)}_{ab}= \delta_{a\, 3}\delta_{b\, 5}  \quad\quad\quad  {\cal Q}^{(8)}_{ab}=  -\delta_{a\, 8}\delta_{b\, 7} ~.
\end{eqnarray}
The only non-vanishing combinations turn out to be
\begin{eqnarray}
\langle {\mathbf{x}}\ a|-g\gamma\ ^{\ast}\! D(A)^{-1}|{\mathbf{y}}\ b\rangle &=&
 \delta_{a b}\delta({\mathbf{x}}-{\mathbf{y}})
+\sum_{n=1}^\infty\langle {\mathbf{x}}\ a| \left(  {\cal Q}^{(0)}\, \partial_1 {1\over X_3}\right)^n|{\mathbf{y}}\ b\rangle\ 
\nonumber\\
&&\!\!\!\!\!\!
+\sum_{n=1}^\infty\langle {\mathbf{x}}\ a| \left({\cal Q}^{(+)}\, \partial_1 {1\over X_+}\right)^n|{\mathbf{y}}\ b\rangle\ 
+\sum_{n=1}^\infty\langle {\mathbf{x}}\ a| \left({\cal Q}^{(-)}\, \partial_1 {1\over X_-}\right)^n|{\mathbf{y}}\ b\rangle\ 
\nonumber\\
&&\!\!\!\!\!\!\!\!\!\!\!\!\!\!\!\!\!\!\!\!\!\!\!\!\!\!\!\!\!\!\!\!\!\!\!\!\!\!\!\!\!\!\!\!\!\!\!\!\!\!\!\!\!\!\!\!\!\!
+\sum_{n=0}^\infty\langle {\mathbf{x}}\ a|  {\cal Q}^{(3)}\, \partial_2 {1\over X_+}
\left( {\cal Q}^{(+)}\, \partial_1 {1\over X_+} \right)^n|{\mathbf{y}}\ b\rangle\ 
+\sum_{n=0}^\infty\langle {\mathbf{x}}\ a|  {\cal Q}^{(8)}\, \partial_2 {1\over X_-}
 \left( {\cal Q}^{(-)}\, \partial_1 {1\over X_-}\right)^n|{\mathbf{y}}\ b\rangle~. 
\end{eqnarray}
This leads to the possibility to write the non-local part of the kernel (\ref{kernel}) in much simpler form.

\subsection{"Weyl-decomposition" of the non-local potential}

Writing the non-local part of the kernel (\ref{kernel}) with two arbitrary functions $ {\cal R}_a({\mathbf{x}})$ 
and $ {\cal O}_a({\mathbf{x}})$,
\begin{eqnarray}
          {\cal R}_a({\mathbf{x}})\,  \langle {\mathbf{x}}\ a|
      ^{\ast}\! D(A)^{-1}\,  ^{\ast}\! D^{\dagger}(A)^{ -1} |{\mathbf{y}}\ b\rangle \, {\cal O}_b({\mathbf{y}}) ~,
\end{eqnarray}
 in the form
\begin{eqnarray}
{1\over g^2}       {\cal R}_a  ({\mathbf{x}})\, \gamma^{-1}_{ac} ({\mathbf{x}})
\langle {\mathbf{x}}\ c|\,\left(- g \gamma\  ^{\ast}\! D(A)^{-1}\right)
\, \, \left(- g \gamma\ ^{\ast}\! D(A)^{ -1}\right)^{\dagger}  |{\mathbf{y}}\ d\rangle \, 
\gamma^{-1\, T}_{db}({\mathbf{y}}) {\cal O}_b({\mathbf{y}}) ~,
\end{eqnarray}
and noting that we can write
\begin{eqnarray}
&&\!\!\!\!\!\!\!\!\!\!\!\!\!\!\!\!\!\!\!\!\!\!\!\!\!\!\!
 {\gamma^{-1\, T}_{1b}{\cal O}_b\choose  \gamma^{-1\, T}_{2b}{\cal O}_b}
= {1\over X_3}Q^T{\widetilde{\cal O}_1 \choose \widetilde{\cal O}_2}~,
\quad\quad
 {\gamma^{-1\, T}_{4b}{\cal O}_b\choose  \gamma^{-1\, T}_{5b}{\cal O}_b}
= - {1\over X_+}Q^T{\widetilde{\cal O}_4 \choose \widetilde{\cal O}_5}~,
\quad\quad
 {\gamma^{-1\, T}_{6b}{\cal O}_b\choose  \gamma^{-1\, T}_{7b}{\cal O}_b}
={1\over X_-}Q^T{\widetilde{\cal O}_6 \choose \widetilde{\cal O}_7}~,
\nonumber\\
&& \gamma^{-1\, T}_{3b}{\cal O}_b = {1\over Y_4} \left( {\cal O}_3 +{1\over\sqrt{3} }{\cal O}_8\right)~,
\quad\quad
\gamma^{-1\, T}_{8b}{\cal O}_b = {1\over Y_6} \left( {\cal O}_3 -{1\over\sqrt{3} }{\cal O}_8\right)~,
\end{eqnarray}
with the two-dimensional matrix
\begin{eqnarray}
Q:= \left(
\begin{array}{c c }
0&1 \\
-1&0 
\end{array}
\right)~,
\end{eqnarray}
and the shifted
\begin{eqnarray}
\widetilde{\cal O}_1 &:=& {\cal O}_1
                               -{Y_6\over 2 Y_4}\left( {\cal O}_3 +{1\over\sqrt{3} }{\cal O}_8\right)
                               +{Y_4\over 2 Y_6}\left( {\cal O}_3 -{1\over\sqrt{3} }{\cal O}_8\right)~,
\nonumber\\
\widetilde{\cal O}_2 &:=&{\cal O}_2~,
\nonumber\\
\widetilde{\cal O}_4 &:=& {\cal O}_4
                               -{Y_1\over 2 Y_6}\left( {\cal O}_3 -{1\over\sqrt{3} }{\cal O}_8\right)
                               -{Y_+\over  Y_4}\left( {\cal O}_3 +{1\over\sqrt{3} }{\cal O}_8\right)~,
\nonumber\\
\widetilde{\cal O}_5 &:=& {\cal O}_5
                               -{Y_2\over 2 Y_6}\left( {\cal O}_3 -{1\over\sqrt{3} }{\cal O}_8\right)~,
\nonumber\\
\widetilde{\cal O}_6 &:=& {\cal O}_6
                               +{Y_1\over 2 Y_4}\left( {\cal O}_3 +{1\over\sqrt{3} }{\cal O}_8\right)
                               -{Y_-\over  Y_6}\left( {\cal O}_3 -{1\over\sqrt{3} }{\cal O}_8\right)~,
\nonumber\\
\widetilde{\cal O}_7 &:=& {\cal O}_7
                               -{Y_2\over 2 Y_4}\left( {\cal O}_3 +{1\over\sqrt{3} }{\cal O}_8\right)~,
\label{cal-O}
\end{eqnarray}
we obtain the decomposition ("Weyl-decomposition") into the four terms
\begin{eqnarray}
 \sum_{a,b=1}^8  {\cal R}_a({\mathbf{x}})\,  \langle {\mathbf{x}}\ a|
      ^{\ast}\! D(A)^{-1} \left(  ^{\ast}\! D(A)^{ -1}\right)^{\dagger} |{\mathbf{y}}\ b\rangle \, {\cal O}_b({\mathbf{y}}) &=&
\nonumber\\
&&\!\!\!\!\!\!\!\!\!\!\!\!\!\!\!\!\!\!\!\!\!\!\!\!\!\!\!\!\!\!\!\!\!\!\!\!\!\!\!\!\!\!\!\!\!\!\!\!\!\!\!
\!\!\!\!\!\!\!\!\!\!\!\!\!\!\!\!\!\!\!\!\!\!\!\!\!\!\!\!\!\!\!\!\!\!\!\!\!\!\!\!\!\!\!\!\!\!\!\!\!\!\!
\!\!\!\!\!\!\!\!\!\!\!\!\!\!\!\!\!\!\!\!\!\!\!\!\!\!\!
 ={\widetilde{\cal R}_1^\dagger ({\mathbf{x}}) \choose\widetilde{\cal R}_2^\dagger ({\mathbf{x}})}^{\!\! T} \,
\langle {\mathbf{x}}|\, ^{\ast}\! D_1(X_3)^{-1}\,\left( ^{\ast}\! D_1(X_3)^{-1}\right)^\dagger |{\mathbf{y}}\rangle
{\widetilde{\cal O}_1({\mathbf{y}}) \choose \widetilde{\cal O}_2({\mathbf{y}})}
\nonumber\\
&&\!\!\!\!\!\!\!\!\!\!\!\!\!\!\!\!\!\!\!\!\!\!\!\!\!\!\!\!\!\!\!\!\!\!\!\!\!\!\!\!\!\!\!\!\!\!\!\!\!\!\!
\!\!\!\!\!\!\!\!\!\!\!\!\!\!\!\!\!\!\!\!\!\!\!\!\!\!\!\!\!\!\!\!\!\!\!\!\!\!\!\!\!\!\!\!\!\!\!\!\!\!\!
\!\!\!\!\!\!\!\!\!\!\!\!\!\!\!\!\!\!\!\!\!\!\!
+{\widetilde{\cal R}_4^\dagger  ({\mathbf{x}}) ,\choose
\widetilde{\cal R}_5^\dagger  ({\mathbf{x}})
 -\partial_2^{ ({\mathbf{x}})}\!\left[\!
\left({\cal R}_3({\mathbf{x}})+{1\over \sqrt{3}}{\cal R}_8({\mathbf{x}})\right)  Y_4^{-1} ({\mathbf{x}})\right]}^{\!\! T}\!\!\times
\nonumber\\
&&\!\!\!\!\!\!\!\!\!\!\!\!\!\!\!\!\!\!\!\!\!\!\!\!\!\!\!\!\!\!\!\!\!\!\!\!\!\!\!\!\!\!\!\!\!\!\!\!\!\!\!
\!\!\!\!\!\!\!\!\!\!\!\!\!\!\!\!\!\!\!\!\!\!\!\!\!\!\!\!\!\!\!\!\!\!\!\!\!\!\!\!\!\!\!\!\!
\!\!\!\!\!\!\!\!\!\!\!\!
\langle {\mathbf{x}}|\, ^{\ast}\! D_1(-X_+)^{-1}\,\left( ^{\ast}\! D_1(-X_+)^{-1}\right)^\dagger |{\mathbf{y}}\rangle
{ \widetilde{\cal O}_4({\mathbf{y}})  \choose \widetilde{\cal O}_5({\mathbf{y}}) 
-\partial_2^{ ({\mathbf{y}})}\!\left[Y_4^{-1}({\mathbf{y}})\left({\cal O}_3({\mathbf{y}})
+{1\over \sqrt{3}}{\cal O}_8({\mathbf{y}})\right)\right]}
\nonumber\\
&&\!\!\!\!\!\!\!\!\!\!\!\!\!\!\!\!\!\!\!\!\!\!\!\!\!\!\!\!\!\!\!\!\!\!\!\!\!\!\!\!\!\!\!\!\!\!\!\!\!\!\!
\!\!\!\!\!\!\!\!\!\!\!\!\!\!\!\!\!\!\!\!\!\!\!\!\!\!\!\!\!\!\!\!\!\!\!\!\!\!\!\!\!\!\!\!\!\!\!\!\!\!\!
\!\!\!\!\!\!\!\!\!\!\!\!\!\!\!\!\!\!\!\!\!\!\!
+{ \widetilde{\cal R}_6^\dagger ({\mathbf{x}}) \choose 
\widetilde{\cal R}_7^\dagger({\mathbf{x}}) 
+\partial_2^{ ({\mathbf{x}})}\left[\left(\!{\cal R}_3 ({\mathbf{x}})-{1\over \sqrt{3}}{\cal R}_8 ({\mathbf{x}})\!\right)
 \!\, Y_6^{-1}({\mathbf{x}})\right]}^{\!\! T}\!\!\times
\nonumber\\
&&\!\!\!\!\!\!\!\!\!\!\!\!\!\!\!\!\!\!\!\!\!\!\!\!\!\!\!\!\!\!\!\!\!\!\!\!\!\!\!\!\!\!\!\!\!\!\!\!\!\!\!
\!\!\!\!\!\!\!\!\!\!\!\!\!\!\!\!\!\!\!\!\!\!\!\!\!\!\!\!\!\!\!\!\!\!\!\!\!\!\!\!\!\!\!\!\!
\!\!\!\!\!\!\!\!\!\!\!\!\
\langle {\mathbf{x}}|\, ^{\ast}\! D_1(X_-)^{-1}\,\left( ^{\ast}\! D_1(X_-)^{-1}\right)^\dagger |{\mathbf{y}}\rangle
{ \widetilde{\cal O}_6({\mathbf{y}})\choose \widetilde{\cal O}_7({\mathbf{y}})
 +\partial_2^{ ({\mathbf{y}})}\left[Y_6^{-1}({\mathbf{y}})\left({\cal O}_3({\mathbf{y}})
-{1\over \sqrt{3}}{\cal O}_8({\mathbf{y}})\right)\!\right]}
\nonumber\\
&&\!\!\!\!\!\!\!\!\!\!\!\!\!\!\!\!\!\!\!\!\!\!\!\!\!\!\!\!\!\!\!\!\!\!\!\!\!\!\!\!\!\!\!\!\!\!\!\!\!\!\!
\!\!\!\!\!\!\!\!\!\!\!\!\!\!\!\!\!\!\!\!\!\!\!\!\!\!\!\!\!\!\!\!\!\!\!\!\!\!\!\!\!\!\!\!\!\!\!\!\!\!\!
\!\!\!\!\!\!\!\!\!\!\!\!\!\!\!\!\!\!\!\!\!\!\!\!
+\!\Bigg[ \left(\! {\cal R}_3 ({\mathbf{x}}) +{1\over\sqrt{3} }{\cal R}_8 ({\mathbf{x}})\!\right)\!  
{1\over Y_4^2({\mathbf{x}})}
\! \left(\! {\cal O}_3 ({\mathbf{x}}) +{1\over\sqrt{3} }{\cal O}_8 ({\mathbf{x}})\!\right)\!
\nonumber\\
&&\!\!\!\!\!\!\!\!\!\!\!\!\!\!\!\!\!\!\!\!\!\!\!\!\!\!\!\!\!\!\!\!\!\!\!\!\!\!\!\!\!\!\!\!\!\!\!\!\!\!\!
\!\!\!\!\!\!\!\!\!\!\!\!\!\!\!\!\!\!\!\!\!\!\!\!\!\!\!\!\!\!\!\!\!\!\!\!\!\!\!\!\!\!\!\!\!\!\!\!\!\!\!
+\left(\! {\cal R}_3({\mathbf{x}}) -{1\over\sqrt{3} }{\cal R}_8({\mathbf{x}})\!\right)\! 
 {1\over Y_6^2({\mathbf{x}})} 
\!\left(\! {\cal O}_3({\mathbf{x}}) -{1\over\sqrt{3} }{\cal O}_8({\mathbf{x}})\!\right)\!\Bigg]
\!\delta({\mathbf{x}}-{\mathbf{y}})~,
\label{Weyl-decomp}
\end{eqnarray}
in terms of the two-dimensional Faddeev-Popov operator
\begin{eqnarray}
\langle {\mathbf{x}}|\,  ^*\! D_1(f)|{\mathbf{y}}\rangle& := &
\left(
\begin{array}{c c }
1&0 \\
0&1 
\end{array}
\right)\partial_1^{({\mathbf{x}})}
     \left[\delta({\mathbf{x}}-{\mathbf{y}})\right]
+g \left(
\begin{array}{c c }
0&1 \\
-1&0 
\end{array}
\right) f({\mathbf{x}})\, \delta({\mathbf{x}}-{\mathbf{y}})
\nonumber\\
&\equiv & 
1_2\, \partial_1^{({\mathbf{x}})}
     \left[\delta({\mathbf{x}}-{\mathbf{y}})\right]
+g\, Q\,  f({\mathbf{x}})\, \delta({\mathbf{x}}-{\mathbf{y}})~.
\label{D-2-dim}
\end{eqnarray}
The inverse of the two-dimensional FP operator (\ref{D-2-dim}) has  the expansion
\begin{eqnarray}
\langle {\mathbf{x}}|\, ^{\ast}\! D_1(f)^{-1}|{\mathbf{y}}\rangle &=&
-{1 \over g}\, Q\, {1\over  f({\mathbf{x}})} \Bigg[1_2\,
\delta({\mathbf{x}}-{\mathbf{y}})+{1 \over g}\, Q^T \,
     \partial_1^{({\mathbf{x}})}
     \left[{1\over f ({\mathbf{x}}) } \delta({\mathbf{x}}-{\mathbf{y}})\right] 
\nonumber\\
&&\quad\quad\quad\quad\quad\quad
-{1 \over g^2}\, 1_2\,
     \partial_1^{({\mathbf{x}})}\!\!\left[{1\over f({\mathbf{x}})}
     \partial_1^{({\mathbf{x}})}\!\!
     \left[{1\over f({\mathbf{x}})}\delta({\mathbf{x}}-{\mathbf{y}})\right]\right]
+...\Bigg]~.
\label{D-2-dim-exp}
\end{eqnarray}
Thus (using $Q^T=-Q$ and $Q^2=-1_2$)  we obtain
\begin{eqnarray}
\!\!\!\!\!\!\!\!\!\!\!\!
\langle {\mathbf{x}}|\, ^{\ast}\! D_1(f)^{-1}\left( ^{\ast}\! D_1(f)^{-1}\right)^{\!\dagger}\! |{\mathbf{y}}\rangle
\!\!\!\! &=&\!\!\!\!
-{1\over g^2} \!\! \sum_{m,n=0}^\infty {(-1)^n\over g^{m+n}}
{1\over  f({\mathbf{x}})}\!
\left(\stackrel{\leftarrow}{\partial}_1^{({\mathbf{x}})} \!\! {1\over  f({\mathbf{x}})} \right)^{\!\! m}
\!\!\delta({\mathbf{x}}-{\mathbf{y}})\!
\left(\!{1\over  f({\mathbf{y}})} \stackrel{\rightarrow}{\partial}_1^{({\mathbf{y}})}\right)^{\!\! n}  \!\!
{1\over  f({\mathbf{y}})}\, Q^{m+n},
\label{Weyl-decomp-expand}
\end{eqnarray}
and hence an explicit derivative expansion of (\ref{Weyl-decomp}).

\section{Expansion of the Hamiltonian in spatial derivatives}

In order to obtain a consistent expansion of the physical Hamiltonian in spatial derivatives,
also the non-locality of the Jacobian ${\cal J}$
must be taken into account. This can be achieved in the following way.

\subsection{Inclusion of the non-locality of the Jacobian into the Hamiltonian}

Rewriting the FP operator as the matrix product
\begin{eqnarray}
^{\ast}\! D_{ab}(A)
    =-g\gamma_{ac}(A)\Big[\delta_{cb}-{1\over g}\gamma^{-1}_{cd}(A)\left( \Gamma_{i}\right)_{db}\partial_i\Big]
    \equiv -g\gamma_{ac}(A)\,^{\ast}\! \widetilde{D}_{cb}(A)~,
\nonumber
\end{eqnarray}
the Jacobian ${\cal J}$ factorizes
\begin{equation}
{\cal J}={\cal J}_0 \widetilde{\cal J}~,
\end{equation}
with the local
\begin{equation}
{\cal J}_0 \equiv\det |\gamma|=\prod_{\mathbf{x}}\det|\gamma({\mathbf{x}})|~, \quad \quad
\det|\gamma({\mathbf{x}})|=X^2_3\left(X^2_3-3 X^{2}_8\right)^2\  Y_4 Y_6 ~,
\end{equation}
and the non-local
$\widetilde{\cal J}\equiv\det |^{\ast}\! \widetilde{D}|$~.
Including furthermore the non-local part of the measure into the wave functional
\begin{eqnarray}
\widetilde{\Psi}(A):=\widetilde{\cal J}^{1/2}\Psi(A)~,
\nonumber
\end{eqnarray}
leads to the transformed Hamiltonian
$
\widetilde{H}:=\widetilde{\cal J}^{1/2} H \widetilde{\cal J}^{-1/2}~,
$
Hermitean with respect to the local measure ${\cal J}_0$
\begin{eqnarray}
\label{Htilde}
\widetilde{H}(A,P)\!\!\!\!&=&
    \!\!\!\! {1\over 2}{\cal J}_0^{-1}\!\!\int d^3{\mathbf{x}}\int d^3{\mathbf{y}}
\ P_{ai}({\mathbf{x}})\ {\cal J}_0\  {\cal K}_{ai|bk}({\mathbf{x}},{\mathbf{y}}) P_{bk}({\mathbf{y}})
     +{1\over 2}\int d^3{\mathbf{x}}\left(B_{ai}(A)\right)^2+V_{\rm meas}(A)~,
\end{eqnarray}
at the cost of extra terms\footnote{
Although $V_{\rm meas}$ is,  in principle, part of the electric term of the
Hamiltonian, I shall treat it separately in this work as the socalled "measure term".
} $V_{\rm meas}$ arising from the non-local factor $\widetilde{\cal J}$
of the original measure ${\cal J}$
\begin{eqnarray}
V_{\rm meas}(A)&=& { \delta({\mathbf{0}})\over 4}{\cal J}_0^{-1}\int d^3{\mathbf{x}}
                       \frac{\delta}{\delta A_{ai}({\mathbf{x}})}\left[{\cal J}_0\int d^3{\mathbf{y}}\ 
                       {\cal K}_{ai|bk}({\mathbf{x}},{\mathbf{y}})\,
                      \Delta_{bk}({\mathbf{y}})\right]
\nonumber\\
&&\quad\quad 
+{ \delta({\mathbf{0}})^2\over 8}\int d^3{\mathbf{x}}\int d^3{\mathbf{y}}\ 
   \Delta_{ai}({\mathbf{x}})\ {\cal K}_{ai|bk}({\mathbf{x}},{\mathbf{y}})\,
  \Delta_{bk}({\mathbf{y}})~,
\label{Vmeas}
\end{eqnarray}
(the regularisation of $ \delta({\mathbf{0}})$ will be given later) with the original kernel ${\cal K}$ in (\ref{kernel}) and the
\begin{eqnarray}
\!\!\!\!\!\!\!\!\!\!\!\!\!
\Delta_{ai}({\mathbf{x}})\!\!\!\!\!&:=&\!\!\!\!\!   \delta({\mathbf{0}})^{-1} 
                    \frac{\delta \ln \widetilde{\cal J}}{\delta A_{ai}({\mathbf{x}})}=
          - \delta({\mathbf{0}})^{-1}\langle {\mathbf{x}}\ b|g\, ^{\ast}\!D^{-1}+\,\gamma^{-1}|{\mathbf{x}}\ c\rangle
             \   {\delta  \gamma_{cb}  \over \delta A_{ai}}\equiv  {\cal U}_{bc}({\mathbf{x}})
\   {\delta  \gamma_{cb}  \over \delta A_{ai}} ~.
\end{eqnarray}
with the  
\begin{eqnarray}
{\cal U} _{ab}({\mathbf{x}}) &=&
-{1 \over g}\gamma^{-1}_{ac}({\mathbf{x}})\left(\Gamma_i\right)_{cd}
     \partial_i
     \gamma^{-1}_{db}({\mathbf{x}})
-{1 \over g^2}\gamma^{-1}_{ac}({\mathbf{x}})\left(\Gamma_i\right)_{cd}
     \partial_i\left[\gamma^{-1}_{dm}({\mathbf{x}})\left(\Gamma_j\right)_{mn}
     \partial_j
    \gamma^{-1}_{nb}({\mathbf{x}})\right]
+...~,
\end{eqnarray}
and the c-numbers $ \delta  \gamma_{cb} / \delta A_{ai}$ obtained from (\ref{gamma}).
It turns out that the only non-vanishing $\Delta$ are
\begin{eqnarray}
\Delta_{31}({\mathbf{x}})  = {\cal U}_{[12]}({\mathbf{x}})
+{1\over 2}\left({\cal U}_{[45]}({\mathbf{x}})-{\cal U}_{[67]}({\mathbf{x}})\right)~,
 &&   
\Delta_{81}({\mathbf{x}})  ={\sqrt{3}\over 2}\left({\cal U}_{[45]}({\mathbf{x}})
+{\cal U}_{[67]}({\mathbf{x}})\right)~,
\end{eqnarray}
being linear combinations of  
\begin{eqnarray}
 {\cal U}_{[12]}= 2\, u[X_3]~,\quad\quad 
 {\cal U}_{[45]}=-2\,  u [X_+]~,\quad\quad
 {\cal U}_{[67]}=2\, u [X_-]~,
\end{eqnarray}
in terms of the functional
\begin{eqnarray}
  u[f]&:=&
-{1 \over g^2} {1\over f}\partial_1\left[ {1\over f} 
     \partial_1 {1\over f} \right]
-{1 \over g^4}  {1\over f}\partial_1\left[ {1\over f}
     \partial_1\left[  {1\over f} \partial_1\left[  {1\over f}
     \partial_1  {1\over f} \right] \right] \right]-...~.
\label{cal-U}
\end{eqnarray}
Hence we have the expansion
\begin{eqnarray}
\Delta_{a1}({\mathbf{x}})&=&\Delta_{a1}^{(\partial_1\partial_1 )}[X({\mathbf{x}})]
+\Delta_{a1}^{(\partial_1\partial_1 \partial_1\partial_1)}[X({\mathbf{x}})]+... \quad\quad\quad a=3,8~,
\end{eqnarray}
containing an even number of spatial derivatives $\partial_1$.

\subsection{Canonical form of the transformed Hamiltonian}

The transformed Hamiltonian (\ref{Htilde}) can be written in the canonical form
\begin{eqnarray}
\label{Htilde2}
\widetilde{H}(A,P)\!\!\!\!&=&
    \!\!\!\! {1\over 2}\int d^3{\mathbf{x}}\ {\cal J}_0^{-1}({\mathbf{x}})
\ P_{ai}({\mathbf{x}})\ {\cal J}_0({\mathbf{x}})\  P_{ai}({\mathbf{x}})
   + V_{\rm elec}(A,P) +V_{\rm magn}(A)+V_{\rm meas}(A)~.
\end{eqnarray}
The electric potential is given by
\begin{eqnarray}
V_{\rm elec}(A,P)&\equiv &
 {1\over 2}{\cal J}_0^{-1}\int d^3{\mathbf{x}}\int d^3{\mathbf{y}}
 \, {\cal G}_a({\mathbf{x}}) \, {\cal J}_0\,  \langle {\mathbf{x}}\ a|
      ^{\ast}\! D(A)^{-1} \left(  ^{\ast}\! D(A)^{ -1}\right)^{\dagger} |{\mathbf{y}}\ b\rangle \,  {\cal G}_b({\mathbf{y}}) ~,
\label{Velec}
\end{eqnarray}
where 
\begin{eqnarray}
{\cal G}_a:= D_i(A)_{ab} P_{bi}=\partial_i P_{ai} +g\, T_a
\end{eqnarray}
in terms of the operators
\begin{equation}
T_a(A,P):=f_{abc}A_{bi}P_{ci}
\equiv T_a^{Y}(Y,P_Y)+T_a^{Z}(Z,P_Z)~.
\label{defT}
\end{equation}
Note, that the components of the (non-reduced) $T_a^{Z} =\! -i f_{abc}\, Z_b\, \partial/\partial Z_c $ 
 satisfy the $su(3)$ algebra
\begin{equation}
[T^{Z}_a,T^{Z}_b]=i\, f_{abc}\, T^{Z}_c~,
\end{equation}
whereas  the reduced $T_a^{Y} =\! -i f_{abc}\, Y_b\, \partial/\partial Y_c $ do not.

The magnetic potential is
\begin{eqnarray}
V_{\rm magn}(A)\equiv {1\over 2}\int d^3{\mathbf{x}}\left(B_{ai}(A)\right)^2~,
\label{Vmagn}
\end{eqnarray}
and the measure terms
\begin{eqnarray}
V_{\rm meas}(A)\equiv V_{\rm meas,I}(A)+ V_{\rm meas,II}(A)~,
\label{Vmeas2}
\end{eqnarray}
are given as
\begin{eqnarray}
V_{\rm meas,I}(A)
&=& { \delta({\mathbf{0}})\over 4}{\cal J}_0^{-1}\int d^3{\mathbf{x}}
                      \left[ \frac{\delta}{\delta X_3({\mathbf{x}})}\left({\cal J}_0\,
                                     \Delta_{31}[{\mathbf{x}}]\right)
                            + \frac{\delta}{\delta X_{8}({\mathbf{x}})}\left({\cal J}_0\,
                                  \Delta_{81}{\mathbf{x}}]\right)\right]
\nonumber\\
&&\!\!\!\!\!\!\!\!
- { \delta({\mathbf{0}})\over 4}{\cal J}_0^{-1}\int d^3{\mathbf{x}}
                       \frac{\delta}{\delta X_{a}({\mathbf{x}})}\Bigg[{\cal J}_0\int d^3{\mathbf{y}}
\, \,  \partial_1^{({\mathbf{x}})} 
  \langle {\mathbf{x}}\ a|
      ^{\ast}\! D(A)^{-1} \left(  ^{\ast}\! D(A)^{ -1}\right)^{\dagger} |{\mathbf{y}}\ b\rangle\ 
 \partial_1^{({\mathbf{y}})}[\Delta_{b1}({\mathbf{y}})]\Bigg]
\nonumber\\
&&\!\!\!\!\!\!\!\!\!\!\!\!\!\!\!\!\!\!\!\!\!\!\!\!\!\!\!\!\!\!
- { \delta({\mathbf{0}})\over 4}{\cal J}_0^{-1}\int d^3{\mathbf{x}}
                       \frac{\delta}{\delta Y_{a}({\mathbf{x}})}\Bigg[{\cal J}_0\int d^3{\mathbf{y}}
\, \,   D_2(Y)_{ac}^{({\mathbf{x}})} 
  \langle {\mathbf{x}}\ c|
      ^{\ast}\! D(A)^{-1} \left(  ^{\ast}\! D(A)^{ -1}\right)^{\dagger} |{\mathbf{y}}\ b\rangle\ 
 \partial_1^{({\mathbf{y}})}[\Delta_{b1}({\mathbf{y}})]\Bigg]~,
\label{VmeasI}
\\
V_{\rm meas, II}(A)
&=&
{ \delta({\mathbf{0}})^2\over 8}\int d^3{\mathbf{x}}\ 
 \left[ \left(\Delta_{31}({\mathbf{x}})\right)^2+\left(\Delta_{81}({\mathbf{x}})\right)^2\right]
\nonumber\\
&& +
{ \delta({\mathbf{0}})^2\over 8}\int d^3{\mathbf{x}}\int d^3{\mathbf{y}}\ 
 \partial_1 [\Delta_{a1}({\mathbf{x}})]\  \langle {\mathbf{x}}\ a|
      ^{\ast}\! D(A)^{-1} \left(  ^{\ast}\! D(A)^{ -1}\right)^{\dagger} |{\mathbf{y}}\ b\rangle\ 
 \partial_1[\Delta_{b1}({\mathbf{y}})]~,
\label{VmeasII}
\end{eqnarray}
obtained from (\ref{Vmeas}) noting
$
 D_i(A)_{ab} \Delta_{b1}\delta_{i1}=\partial_1 \Delta_{a1}~.
$

Using the Wey-decomposition (\ref{Weyl-decomp}) and the expansion
(\ref{Weyl-decomp-expand}) of the corresponding two-dimensional non-local potentials, we can
further simplify the electric potential (\ref{Velec}) and the measure terms (\ref{VmeasI}) and (\ref{VmeasII}).

\subsection{Weyl-decomposition of the electric potential}

Using (\ref{Weyl-decomp}), the electric potential (\ref{Velec}) can be written in the form
\begin{eqnarray}
V_{\rm elec}(A,P)&= &
{1\over 2}{\cal J}_0^{-1} 
\int d^3{\mathbf{x}}\int d^3{\mathbf{y}}
\Bigg\{
{\widetilde{\cal G}_1^\dagger ({\mathbf{x}}) \choose \widetilde{\cal G}_2^\dagger ({\mathbf{x}})}^{\!\! T} {\cal J}_0\,
\langle {\mathbf{x}}|\, ^{\ast}\! D_1(X_3)^{-1}\,\left( ^{\ast}\! D_1(X_3)^{-1}\right)^\dagger |{\mathbf{y}}\rangle
{\widetilde{\cal G}_1({\mathbf{y}}) \choose \widetilde{\cal G}_2({\mathbf{y}})}
\nonumber\\
&&\quad\quad\quad
+{\widetilde{\cal G}_4^\dagger  ({\mathbf{x}}) \choose 
\widetilde{\cal G}_5^\dagger  ({\mathbf{x}})
 -\partial_2^{ ({\mathbf{x}})}\!\left[\!
\left({\cal G}_3 ({\mathbf{x}})+{1\over \sqrt{3}}{\cal G}_8 ({\mathbf{x}})\right) Y_4^{-1} ({\mathbf{x}})\right]}^{\!\! T}
 {\cal J}_0\, \times
\nonumber\\
&&\langle {\mathbf{x}}|\, ^{\ast}\! D_1(-X_+)^{-1}\,\left( ^{\ast}\! D_1(-X_+)^{-1}\right)^\dagger |{\mathbf{y}}\rangle
{ \widetilde{\cal G}_4({\mathbf{y}})  \choose \widetilde{\cal G}_5({\mathbf{y}}) 
-\partial_2^{({\mathbf{y}})}\!\left[Y_4^{-1}({\mathbf{y}})\left({\cal G}_3({\mathbf{y}})+{1\over \sqrt{3}}
{\cal G}_8({\mathbf{y}})\right)\right]}
\nonumber\\
&&\quad\quad\quad
+{ \widetilde{\cal G}_6^\dagger ({\mathbf{x}}) \choose 
\widetilde{\cal G}_7^\dagger({\mathbf{x}}) 
+\partial_2^{ ({\mathbf{x}})}\left[\left(\!{\cal G}_3 ({\mathbf{x}})-{1\over \sqrt{3}}
{\cal G}_8 ({\mathbf{x}}) \!\right)\, Y_6^{-1}({\mathbf{x}})\right]
}^{\!\! T} {\cal J}_0\, \times
\nonumber\\
&&\langle {\mathbf{x}}|\, ^{\ast}\! D_1(X_-)^{-1}\,\left( ^{\ast}\! D_1(X_-)^{-1}\right)^\dagger |
{\mathbf{y}}\rangle
{ \widetilde{\cal G}_6({\mathbf{y}})\choose \widetilde{\cal G}_7({\mathbf{y}})
 +\partial_2^{ ({\mathbf{y}})}\left[Y_6^{-1}({\mathbf{y}})\left({\cal G}_3({\mathbf{y}})
-{1\over \sqrt{3}}{\cal G}_8({\mathbf{y}})\right)\!\right]}
\Bigg\}
\nonumber\\
&&
+{1\over 2}{\cal J}_0^{-1} 
\int d^3{\mathbf{x}}
\Bigg\{\! \left(\! {\cal G}_3({\mathbf{x}})  +{1\over\sqrt{3} }{\cal G}_8({\mathbf{x}}) \!\right)   {\cal J}_0\, 
{1\over Y_4^2({\mathbf{x}})}
\! \left(\! {\cal G}_3({\mathbf{x}})   +{1\over\sqrt{3} }{\cal G}_8({\mathbf{x}}) \!\right)\!
\nonumber\\
&&\quad\quad\quad\quad\quad\quad\quad\quad\quad\quad
+\left(\! {\cal G}_3({\mathbf{x}}) -{1\over\sqrt{3} }{\cal G}_8({\mathbf{x}}) \!\right) {\cal J}_0\, 
 {1\over Y_6^2({\mathbf{x}})} 
\!\left(\! {\cal G}_3({\mathbf{x}}) -{1\over\sqrt{3} }{\cal G}_8({\mathbf{x}})\!\right)  \Bigg\}~,
\label{elecWD}
\end{eqnarray}
in terms of the two-dimensional Faddeev-Popov operator (\ref{D-2-dim})
and the shifted $\widetilde{\cal G}_a$ defined according to (\ref{cal-O}).
Using (\ref{D-2-dim-exp}) we can expand the electric potential (\ref{elecWD}) as
\begin{eqnarray}
\label{expand}
V_{\rm elec}&=&
  {1\over 2\, g^2}{\cal J}_0^{-1}\sum_{m,n=0}^\infty {(-1)^n\over  g^{m+n}}\int d^3{\mathbf{x}}
\Bigg\{
 {\widetilde{\cal G}_1^\dagger \choose \widetilde{\cal G}_2^\dagger }^{\!\! T}
{1\over  X_3}
\left(\stackrel{\leftarrow}{\partial}_1  {1\over  X_3} \right)^{\!\! m}
\, {\cal J}_0\,
\left({1\over  X_3} \stackrel{\rightarrow}{\partial}_1\right)^{\!\! n} 
{1\over  X_3} Q^{m+n}
{\widetilde{\cal G}_1 \choose \widetilde{\cal G}_2}
\nonumber\\
&&\quad\quad\quad\quad\quad\quad\quad\quad
+{\widetilde{\cal G}_4^\dagger  \choose
\widetilde{\cal G}_5^\dagger 
 -\partial_2\!\left[\!
\left({\cal G}_3+{1\over \sqrt{3}}{\cal G}_8 \right) Y_4^{-1}\right]}^{\!\! T}\times
\nonumber\\
&&\quad\quad\quad
{1\over  X_+}
\left(-\stackrel{\leftarrow}{\partial}_1  {1\over  X_+} \right)^{\!\! m}
\, {\cal J}_0\, 
\left(-{1\over  X_+} \stackrel{\rightarrow}{\partial}_1\right)^{\!\! n} 
{1\over  X_+}\, Q^{m+n}
{ \widetilde{\cal G}_4  \choose \widetilde{\cal G}_5 
-\partial_2\!\left[Y_4^{-1}\left({\cal G}_3+{1\over \sqrt{3}}{\cal G}_8\right)\right]}
\nonumber\\
&&\quad\quad\quad\quad\quad\quad\quad\quad
+{ \widetilde{\cal G}_6^\dagger \choose 
\widetilde{\cal G}_7^\dagger 
+\partial_2\left[\left(\!{\cal G}_3-{1\over \sqrt{3}}{\cal G}_8\!\right)\, Y_6^{-1}\right]}^{\!\! T} 
\times
\nonumber\\
&&\quad\quad\quad
{1\over  X_-}
\left(\stackrel{\leftarrow}{\partial}_1  {1\over  X_-} \right)^{\!\! m}
\, {\cal J}_0\, 
\left({1\over  X_-} \stackrel{\rightarrow}{\partial}_1\right)^{\!\! n} 
{1\over  X_-}Q^{m+n}
{ \widetilde{\cal G}_6\choose \widetilde{\cal G}_7
 +\partial_2\left[Y_6^{-1}\left({\cal G}_3-{1\over \sqrt{3}}{\cal G}_8\right)\right]}
\Bigg\}
\nonumber\\
&& \!\! \!\! \!\!
+ {1\over 2}\int d^3{\mathbf{x}}\Bigg\{ \!\!
\left({\cal G}_3+{1\over \sqrt{3}}{\cal G}_8\right) \!\, {\cal J}_0\,  {1\over Y_4^2}\left({\cal G}_3+{1\over \sqrt{3}}{\cal G}_8\right)
+\left({\cal G}_3-{1\over \sqrt{3}}{\cal G}_8\right) \!\, {\cal J}_0\, {1\over Y_6^2}\left({\cal G}_3-{1\over \sqrt{3}}{\cal G}_8\right)
\!\!\Bigg\}~.
\label{elecWDexp}
\end{eqnarray}

\subsection{Weyl-decomposition of the non-local part of the measure term}

The corresponding expressions for the measure terms,  using the Weyl-decomposition (\ref{Weyl-decomp}), can be obtained in an analogous way.
The first summand of $V_{\rm meas,I}(A)$,
reads
\begin{eqnarray}
V_{\rm meas,I}(A)
&=& { \delta({\mathbf{0}})\over 4}\int d^3{\mathbf{x}}
                      \Big[\left({2\over X_3}-{1\over X_+}-{1\over X_-}+ \frac{\delta}{\delta X_3}\right) u[X_3]+
\left({2\over X_+}-{1\over X_3}-{1\over X_-}+ \frac{\delta}{\delta X_+}\right) u[X_+]
\nonumber\\
&&\quad\quad\quad\quad\quad\quad
+\left({2\over X_-}-{1\over X_+}-{1\over X_3}+ \frac{\delta}{\delta X_-}\right) u[X_-]
                            \Big]+O(\partial^4)~,
\label{VmeasIpart}
\end{eqnarray}
 using the $u$-functionals defined in (\ref{cal-U}) and 
\begin{eqnarray}
 {\cal J}_0^{-1} \frac{\delta}{\delta X_+} {\cal J}_0 =-{2\over X_3}+{2\over X_+}+ \frac{\delta}{\delta X_+}~,  
\quad\quad\quad
 {\cal J}_0^{-1} \frac{\delta}{\delta X_-} {\cal J}_0 =-{2\over X_3}+{2\over X_-}+ \frac{\delta}{\delta X_-}~,  
\end{eqnarray}
and is at least of order $O(\partial^2)$ and the second summand at least of order $O(\partial^4)$.

Similarly, the first summand of $V_{\rm meas,II}(A)$ reads
\begin{eqnarray}
\!\!\!\!\!\!\!\! V_{\rm meas,II}(A)\!\! &=&\!\!
{ \delta({\mathbf{0}})^2\over 4}\!\!\int d^3{\mathbf{x}} 
 \left[ \left(u[X_3]- u[X_+]\right)^2+ \left(u[X_3]-u[X_-]\right)^2+ \left(u[X_+]-u[X_-]\right)^2\right]
+O(\partial^6)~,
\end{eqnarray}
 is at least of order $O(\partial^4)$ and the second summand at least of order $O(\partial^6)$.

\subsection{Derivative expansion of the transformed Hamiltonian and local measure}

Altogether, we obtain an expansion of the transformed physical Hamiltonian (\ref{Htilde2})  in the
number of spatial derivatives
\begin{equation}
\tilde{H} = H_0 + \sum_\alpha V^{(\partial)}_\alpha+
\sum_\beta V^{(\partial\partial)}_\beta
+ ...~,
\end{equation}
with the free part $H_0$ containing no spatial derivatives, the interaction parts
$V^{(\partial)}_\alpha$ containing one spatial derivative, and
$V^{(\partial\partial)}_\beta$ containing two spatial derivatives,
and so on.

The matrix element (\ref{ME}) of a physical operator $O$ becomes the product of local matrix elements
\begin{equation}
\langle \Psi_1| O|\Psi_2\rangle\
\propto
 \prod_{\mathbf{x}}\Big[\int dX({\mathbf{x}})\ X^2_3\left(X^2_3-3\,X^{2}_8\right)^2 \int dY({\mathbf{x}})
\ Y_4 Y_6 \int dZ({\mathbf{x}})
\Big]\ 
\Psi_1^*[X,Y,Z]\ O\ \Psi_2[X,Y,Z]~.
\end{equation}


\section{The free part $H_0$}

The free part $H_0$ containing no spatial derivatives takes the form
\begin{eqnarray}
 H_0  
&=& \int d^3{\mathbf{x}}\, {1\over 2}\Bigg[{\cal J}_0^{-1}\!\!
\ P_{ai}\ {\cal J}_0\ P_{ai}+ {\cal J}_0^{-1}\!\!
\ T_{a}\ {\cal J}_0\left(\gamma^{-1}\gamma^{-1T}\right)_{ac} T_{c}
     +\left(B_{ai}^{\rm hom}(A)\right)^2\Bigg]= \int d^3{\mathbf{x}}\, H_0({\mathbf{x}})~,
\end{eqnarray}
using the homogeneus part $ B^{\rm hom}_{a\, i}:= (1/2) g \epsilon_{ijk}\, f_{abc}\, A_{b\, j}A_{c\, k}  $ 
of the chromomagnetic field 
and the operators $T_a(A,P)$ defined in (\ref{defT}).

Changing to the more convenient polar coordinates
\begin{equation}
X\equiv 
 \left(
\begin{array}{c }
 0   \\ 
 0  \\ 
 r_X  \cos{\psi_X}  \\ 
 0  \\ 
 0 \\ 
 0  \\ 
 0   \\ 
  r_X  \sin{\psi_X}      
\end{array}
\right) \quad\quad\quad
Y\equiv 
 \left(
\begin{array}{c }
  r_{1Y}\cos\theta_Y  \\ 
  r_{1Y}\sin\theta_Y \\ 
   r_Y \cos{\psi_Y} \\ 
   r_{2Y} \\ 
  0 \\ 
   r_{3Y} \\ 
  0  \\ 
   r_Y \sin{\psi_Y}     
\end{array}
\right)
\label{pol coord}
 \end{equation}
and gathering together the parts of the kinetic part of the Hamiltonian depending only on one of the three space 
directions respectively,
we can write 
\begin{eqnarray}
H_0[A({\mathbf{x}}),P({\mathbf{x}})]&=&K_X+K_Y+K_Z+
\nonumber\\
&&
+{1\over 2}\Bigg[\ {1\over  r_X^2 \cos^2{\psi_X}}
\sum_{m=1,2}
\left( {1\over r_{2Y} r_{3Y}}\widetilde{T}_m^{Y\dagger} r_{2Y}\, r_{3Y}+\widetilde{T}_m^{Z}\right)
\left(\widetilde{T}^{Y}_m +\widetilde{T}_m^{Z}\right)
\nonumber\\
&&\quad\quad\quad
+\ {1\over  r_X^2 \cos^2{[\psi_X+2\pi/3]} }
\sum_{m=4,5}
\left( {1\over r_{2Y} r_{3Y}}\widetilde{T}_m^{Y\dagger}r_{2Y}\, r_{3Y}+\widetilde{T}_m^{Z}\right)
\left(\widetilde{T}^{Y}_m +\widetilde{T}_m^{Z}\right)
\nonumber\\
&&\quad\quad\quad\quad\quad
+\ {1\over  r_X^2 \cos^2{[\psi_X+4\pi/3]} }
\sum_{m=6,7}
\left( {1\over r_{2Y} r_{3Y}}\widetilde{T}_m^{Y\dagger}r_{2Y}\,  r_{3Y}+\widetilde{T}_m^{Z}\right)
\left(\widetilde{T}^{Y}_m +\widetilde{T}_m^{Z}\right)
\Bigg]
\nonumber\\
&&
+{1\over 2 r_{2Y}^2 }\left[\left( T^{Z}_3 + {1\over \sqrt{3}}T^{Z}_8 \right)-2i\, {\partial\over  \partial \theta_y}
\right]\left( T^{Z}_3 + {1\over \sqrt{3}}T^{Z}_8 \right)
\nonumber\\
 &&\quad
+{1\over 2  r_{3Y}^2}\left[\left( T^{Z}_3 - {1\over \sqrt{3}}T^{Z}_8 \right)-2i\,  {\partial\over  \partial \theta_y}
\right]\left( T^{Z}_3 - {1\over \sqrt{3}}T^{Z}_8 \right)
+ {1\over 2}\left(B^{\rm hom}_{ai}[X,Y,Z]\right)^2~,
\label{H_0}
\end{eqnarray}
 (noting $ T^{Y}_3= -i\partial/ \theta_Y$ and $ T^{Y}_8=0\, $) with the single-direction kinetic terms
\begin{eqnarray}
\!\!\!\!\!\!\!\!
K_X \!\!\!\! &= & \!\!\!\!
-{1 \over 2 }\left[{\partial^2\over\partial r_X^2}+ { 7\over r_X} {\partial\over\partial r_X}+
 {1 \over r_X^2  } \left(-6 \tan[3 \psi_X]{\partial\over\partial \psi_X}+  {\partial^2\over\partial\psi_X^2}\right)\right]~,
\\
\!\!\!\!\!\!\!\!
K_Y \!\!\!\!  &=& \!\!\!\!
-{1 \over 2 }\Bigg[\left({\partial^2\over\partial r_Y^2} +{ 1\over r_Y} {\partial\over\partial r_Y} +{1\over r_Y^2} 
{\partial^2\over\partial\psi_Y^2}
  \right)
+\sum_{i=1}^3 \left({\partial^2\over\partial r_{iY}^2} +{ 1\over r_{iY} } {\partial\over\partial r_{iY}}
+{1\over   r_{iY}^2} {\partial^2\over\partial\theta_Y^2}\right) 
\Bigg]~,
 \\
\!\!\!\!\!\!\!\!
K_Z \!\!\!\! &=& \!\!\!\!
-{1 \over 2 }\sum_{a=1}^8\left( {\partial\over  \partial Z_{a}}{\partial\over  \partial Z_{a}} \right)~,
\end{eqnarray} 
and the shifted non-Hermitean $\widetilde{T}^{Y}_a$  and Hermitean $\widetilde{T}^{Z}_a$, 
defined according to formulae (\ref{cal-O}).
 
The local part of the Faddeev-Popov determinant, $ {\cal J}_0$,  becomes
\begin{equation}
   {\cal J}_0 =\prod_{\mathbf{x}} {\cal J}_0 ({\mathbf{x}})~, \quad\quad  
 {\cal J}_0 ({\mathbf{x}}) = r_X^{\, 6} ({\mathbf{x}})  \cos^2[3\,\psi_X ({\mathbf{x}})]  
\  r_{2Y} ({\mathbf{x}})\, r_{3Y}  ({\mathbf{x}})~,
\end{equation}
Together with the Jacobians of the coordinate trafo (\ref{pol coord}) the measures
in matrix elements become
\begin{equation}
\langle\Psi_1|{\cal O} |\Psi_2\rangle=\prod_{{\mathbf{x}}}
\int d\mu_X({\mathbf{x}})\int d\mu_Y({\mathbf{x}})\int d\mu_Z({\mathbf{x}})  \ \Psi_1^{\dagger} O\ \Psi_2 ~,
\label{qm-measure}
\end{equation}
with
\begin{eqnarray}
\int d\mu_X &\propto & \int_0^{\infty} d r_X\, r_X^{\, 7}  \int_0^{2\pi} d\psi_X\, \cos^2[3\,\psi_X]~, 
\nonumber\\
\int d\mu_Y &\propto& \int_0^{\infty} d r_Y \, r_Y  \int_0^{2\pi} d\psi_Y \ 
\int_0^{\infty} d  r_{1Y}\,  r_{1Y } \int_0^{\infty} d  r_{2Y}\,  r_{2Y}\int_0^{\infty} d  r_{3Y}\,  r_{3Y}
\int_0^{2\pi} d\theta_Y~,
\nonumber\\
\int d\mu_Z &\propto& \prod_{a=1}^8\int_{-\infty}^{\infty} d Z_a~.
\label{qm-measure2}
\end{eqnarray}
In the space-1 direction the measure has a periodic structure with six zeros at $\psi_X=(2n+1)\pi/6 ~, (n=0,1,...,5)$,
which are Gribov horizons  separating six Weyl-chambers. For recent discussions on Gribov horizons see e.g. \cite{Salgado}.
In the space-3 direction the measure is flat.

\section{First and second order interaction terms}

The interaction parts of first and second order in the number of spatial derivatives, are the following:

\subsection{Magnetic terms}

The first order magnetic part reads
\begin{eqnarray}
\label{Vmagn1}
V^{(\partial)}_{\rm magn}&=&
g \int d^3{\mathbf{x}}\, 
 f_{abc}\,  A_{bj} A_{ck} \partial_j A_{ak}~,
\end{eqnarray}
and the second order magnetic part
\begin{eqnarray}
\label{Vmagn2}
 V^{(\partial\partial)}_{\rm magn} & =&  {1\over 2}\int\!\! d^3{\mathbf{x}}
          \left[ \partial_j A_{ak} \partial_j A_{ak}- \partial_j A_{ak} \partial_k A_{aj}\right]~.
\end{eqnarray}
More explicitly, the first order magnetic parts read
\begin{eqnarray}
V^{(\partial_1)}_{\rm magn}&=&
g \int d^3{\mathbf{x}}\, \Big[
\left( f_{abc}\,  A_{b1} A_{c2}\right) \partial_1 A_{a2}+\left( f_{abc}\,  A_{b1} A_{c3}\right) \partial_1 A_{a3}\Big]~,
\quad 
V^{(\partial_2)}_{\rm magn},V^{(\partial_3)}_{\rm magn}\ \ {\rm cycl.\ perm.}~,
\end{eqnarray}
and the second order magnetic parts
\begin{eqnarray}
 V^{(\partial_1\partial_1)}_{\rm magn} & =&  {1\over 2}\int\!\! d^3{\mathbf{x}}
          \left[ \left(\partial_1 A_{a2}\right)^2+\left(\partial_1 A_{a3}\right)^2\right]~,
\quad\quad 
V^{(\partial_2\partial_2)}_{\rm magn},V^{(\partial_3\partial_3)}_{\rm magn}\ \ {\rm cycl.\ perm.}~,
\\
 V^{(\partial_2\partial_3)}_{\rm magn} & =&  -\int\!\! d^3{\mathbf{x}}
           \left(\partial_2 A_{a3}\right)\left(\partial_3 A_{a2}\right)~,
\quad\quad 
V^{(\partial_1\partial_3)}_{\rm magn},V^{(\partial_1\partial_2)}_{\rm magn}\ \ {\rm cycl.\ perm.}~.
\end{eqnarray}
Note, that there are no higher order magnetic terms.

\subsection{Electric terms}

Up to second order perturbation theory, only
the first order electric parts  $V^{(\partial_1)}_{\rm elec}$, $V^{(\partial_2)}_{\rm elec}$, and
$V^{(\partial_3)}_{\rm elec}$,
and the second order electric parts  $V^{(\partial_1\partial_1)}_{\rm elec}$, $V^{(\partial_2\partial_2)}_{\rm elec}$, and
$V^{(\partial_3\partial_3)}_{\rm elec}$, are needed. They can be easily read off from the general expression (\ref{elecWDexp}). 
The simplest ones are
\begin{eqnarray}
V^{(\partial_3)}_{\rm elec}  
&=& {1\over 2g} \int d^3{\mathbf{x}}\,\Bigg\{
 {1\over  r_X^2 \cos^2{\psi_X}}
\sum_{m=1,2}
\left( {1\over r_{2Y} r_{3Y}} 
(\widetilde{T}_m^{Y\dagger}+\widetilde{T}_m^{Z})\, r_{2Y}\, r_{3Y}\, \widetilde{\partial_3 P_m^{Z}}+{\rm h.c.}
\right)
\nonumber\\
&&\quad\quad\quad\quad\quad
+ {1\over  r_X^2 \cos^2{[\psi_X+2\pi/3]} }
\sum_{m=4,5}
\left(  {1\over r_{2Y} r_{3Y}}
(\widetilde{T}_m^{Y\dagger}+\widetilde{T}_m^{Z})\,  r_{2Y}\, r_{3Y}\, \widetilde{\partial_3 P_m^{Z}}+{\rm h.c.}
\right)
\nonumber\\
&&\quad\quad\quad\quad\quad
+{1\over  r_X^2 \cos^2{[\psi_X+4\pi/3]} }
\sum_{m=6,7}
\left(  {1\over r_{2Y} r_{3Y}}
(\widetilde{T}_m^{Y\dagger}+\widetilde{T}_m^{Z})\,  r_{2Y}\, r_{3Y}\, \widetilde{\partial_3 P_m^{Z}}+{\rm h.c.}
\right)
 \Bigg\}
\nonumber\\
&&\!\!\!\!\!\!\!\!\!\!\!\!\!\!\!\!
- {1\over 2g}\! \int\!\! d^3{\mathbf{x}}\, \Bigg\{
{1\over  r_{2Y}^2 }\!\left( T^Y_3 + T^Z_3 + {1\over \sqrt{3}}T^Z_8 \right)\!\left(\partial_3 P^{Z}_+\right)
+{1\over   r_{3Y}^2}\!\left( T^Y_3 + T^Z_3 - {1\over \sqrt{3}}T^Z_8 \right)\!\left(\partial_3 P^{Z}_-\right)
+{\rm h.c.}\!
 \Bigg\}
~,
\end{eqnarray}
and
\begin{eqnarray}
V^{(\partial_3\partial_3)}_{\rm elec}  
&=&
 {1\over 2g^2} \int d^3{\mathbf{x}}\, \Bigg\{
   {1\over  r_X^2 \cos^2{\psi_X}}\sum_{a=1,2} \left( \widetilde{\partial_3 P_{a}^{Z}}\right)^2
+ {1\over  r_X^2 \cos^2{[\psi_X+2\pi/3]} } \sum_{a=4,5} \left(\widetilde{\partial_3 P_{a}^{Z}}\right)^2
\nonumber\\
&&\quad\quad\quad\quad\quad
+  {1\over  r_X^2 \cos^2{[\psi_X+4\pi/3]} }\sum_{a=6,7} \left( \widetilde{\partial_3 P_{a}^{Z}}\right)^2
+ {1\over r_{2Y}^2} \left(\partial_3 P_{+}^Z\right)^2
+ {1\over r_{3Y}^2} \left(\partial_3 P_{-}^Z\right)^2\Bigg\} .
\end{eqnarray}
The remaining ones are given by somewhat longer expressions. 

\subsection{Measure terms}

Since up to second order in the number of spatial derivatives only 
$\Delta_{31}^{(\partial_1\partial_1 )}$ and $\Delta_{81}^{(\partial_1\partial_1 )}$
are non-vanishing,
the first order measure parts are vanishing
\begin{eqnarray}
 V^{(\partial_1)}_{\rm meas}=0~,\quad\quad
 V^{(\partial_2)}_{\rm meas}=0~,\quad\quad
 V^{(\partial_3)}_{\rm meas}=0~,
\end{eqnarray}
and the only non-vanishing second order measure part is given by the lowest order of
(\ref{VmeasIpart}) obtained using the lowest order $u^{(\partial_1\partial_1)}$ of the $u$-functionals defined in (\ref{cal-U}),
such that 
\begin{eqnarray}
V_{\rm meas}^{(\partial_1\partial_1)}(A)
&=&- { \delta({\mathbf{0}})\over 4 g^2}\int d^3{\mathbf{x}}
                      \Bigg\{\left({2\over X_3}-{1\over X_+}-{1\over X_-}+ \frac{\delta}{\delta X_3}\right)
\left({1\over X_3}\partial_1\left[{1\over X_3}
     \partial_1\left[ {1\over X_3}\right] \right]\right)
\nonumber\\
&&\quad\quad\quad\quad\quad\quad
+\left({2\over X_+}-{1\over X_3}-{1\over X_-}+ \frac{\delta}{\delta X_+}\right)\left({1\over X_+}\partial_1\left[{1\over X_+}
     \partial_1\left[ {1\over X_+}\right] \right]\right)
\nonumber\\
&&\quad\quad\quad\quad\quad\quad
+\left({2\over X_-}-{1\over X_+}-{1\over X_3}+ \frac{\delta}{\delta X_-}\right)\left({1\over X_-}\partial_1\left[{1\over X_-}
     \partial_1\left[ {1\over X_-}\right] \right]\right)
                            \Bigg\}~.
\end{eqnarray}
More explicitly, we have
\begin{eqnarray}
V^{(\partial_1\partial_1)}_{\rm meas} 
&=& 
- { \delta({\mathbf{0}})^2\over 2g^2}\int d^3{\mathbf{x}}\, \Bigg\{ \Bigg[ 
{1\over r_X^2} \left( {1\over \cos^2{[\psi_X]}}+ {1\over \cos{[\psi_X+2\pi/3]}\cos{[\psi_X+4\pi/3]}} \right)
\left(\partial_1 {1\over r_X \cos{[\psi_X]}}\right)^2
\nonumber\\
&&\quad\quad\quad\quad\quad\quad\quad\quad
+\ \ (\psi_X \longrightarrow \psi_X+2\pi/3) \ \ +\ \ (\psi_X \longrightarrow \psi_X+4\pi/3)
\Bigg]
\nonumber\\
&&\quad\quad\quad\quad \quad\quad\quad
+12\left(\partial_1 {1\over r_X^3 \cos{[3\psi_X]}}\right)\left(\partial_1 {1\over r_X \cos{[3\psi_X]}}\right)
\Bigg\}~.
\end{eqnarray}

\subsection{General form of the interaction terms}

The interaction parts of first  order in the number of spatial derivatives
can be written in the general form
\begin{eqnarray}
\label{partial}
V^{(\partial_s)}_{\alpha} &\equiv &\int d {\mathbf{x}}\
\Bigg[\ \widetilde{\cal Y}_{\alpha}[A({\mathbf{x}})]\ \partial_s{\cal Y}_{\alpha}[A({\mathbf{x}})]\ + {\rm h.c.}\Bigg]~,
\quad  s=1,2,3~,
\end{eqnarray}
and those of second order in the number of spatial derivatives
\begin{eqnarray}
\label{partial2}
V^{(\partial_s\partial_t)}_\beta &\equiv & \int d {\mathbf{x}}\ 
\Bigg[
\Big(\partial_s {\cal X}_{\beta}[A({\mathbf{x}})]\Big)
 \widetilde{X}_{\beta}[A({\mathbf{x}})]\,
\left(\partial_t {\cal X}^{\prime }_{\beta}[A({\mathbf{x}})]\right)\ + {\rm h.c.}\Bigg]~,\quad  s,t=1,2,3~,
\end{eqnarray}
containing only first oder
derivatives of functions $\partial_s {\cal X},\partial_s {\cal X}^{\prime }, \partial_s{\cal Y}$.

Interaction parts containing three or four spatial derivatives, 
contain also $\partial_s^2 {\cal X}$and $\partial_s \partial_t {\cal X}$.
For fifths and sixth order we need also the third derivative $\partial^3_s{\cal X}$, and so on.

\section{Coarse graining and strong coupling expansion in $\lambda=g^{-2/3}$}

In order to make the functional approach used above well defined (e.g. what is meant by $\delta({\bf 0})$), 
we should introduce an ultraviolet cutoff which should be rather large
in order to have only slightly spatially varying physical fields for which an expansian in the number of spatial derivatives
is applicable and the flux-tube gauge well-defined.
As for the $SU(2)$-case using the symmetric gauge in \cite{pavel2010}, I shall apply the coarse graining approach
and set an ultraviolet cutoff $a$ by introducing an infinite hypothetical
spatial lattice of granulas
$G({\mathbf{n}},a)$, here large cubes of length $a$,
situated at sites ${\mathbf{x}}=a {\mathbf{n}}$  $({\mathbf{n}}=(n_1,n_2,n_3)\in Z^3)$, and
considering the averaged variables
\begin{equation}
\label{average}
A({\mathbf{n}}) :=  \frac{1}{a^3}\int_{G({\mathbf{n}},a)} d{\mathbf{x}}\ A({\mathbf{x}})
\nonumber
\end{equation}
(where in particular $\delta({\bf 0})\rightarrow 1/a^3$).
Furthermore, the discretised spatial derivatives $\partial_s A({\mathbf{n}}),\partial_s^2 A({\mathbf{n}}) ,...$  
at site ${\mathbf{n}}$ and in the direction $s=1,2,3$ are defined as 
\begin{eqnarray}
\label{1stdisc}
\partial_s A({\mathbf{n}})& := &\lim_{N\rightarrow\infty}\sum_{m=1}^N w_N(m)
\frac{1}{2ma}\left(A({\mathbf{n}}+m {\mathbf{e}}_s)
     -A({\mathbf{n}}-m{\mathbf{e}}_s)\right)~,
\nonumber\\
\partial_s^2 A({\mathbf{n}}) & := &\lim_{N\rightarrow\infty}\sum_{m=1}^N w_N(m)
\frac{1}{(m a)^2 }
\Big(A({\mathbf{n}}+m {\mathbf{e}}_s)
     +A({\mathbf{n}}-m {\mathbf{e}}_s)
     -2 A({\mathbf{n}})\Big)
\nonumber\\
...\quad ...   && ...\quad ... 
\label{discder}
\end{eqnarray}
with the lattice unit vectors ${\mathbf{e}}_1=(1,0,0),{\mathbf{e}}_2=(0,1,0),{\mathbf{e}}_3=(0,0,1)$  
and the distribution  
\begin{equation}
\label{distribution}
w_N(m):=2\frac{(-1)^{m+1}(N!)^2}{(N-m)!(N+m)!}~,\quad 1\leq m\leq N~,\quad\sum_{m=1}^N w_N(m)= 1~.
\end{equation}
The values of $\partial_s A ({\mathbf{n}}), \partial_s^2 A ({\mathbf{n}}),...$ in (\ref{discder})
 for a given site ${\mathbf{n}}$ and direction, say $s=1$,  are chosen
to coincide with the corresponding derivatives
$I'_{2N}(a n_1)|_{n_2,n_3}$, $ I''_{2N}(a n_1)|_{n_2,n_3}$,...\footnote{
Taking the first, second, ... derivatives of the Lagrange interpolation
polynomials $I_{2N}(x)$, with given values
$y_n$ at the equidistant points $x_n=x_0+ n a$, ($n=-N,-N+1,..,N-1,N$),
 at the central point $x_0$, yields: $I'_{2N}(x_0)=\sum_{n=1}^N w_N(n)(y_{n}-y_{-n})/(2na)$,
$I''_{2N}(x_0)=\sum_{n=1}^N w_N(n)(y_{n}+y_{-n}-2y_0)/(na)^2$,..., with the distribution (\ref{distribution}).
The case N=1, in particular, yields $I'_{2}(x_0)=(y_{1}-y_{-1})/(2a)$ and
$I''_{2}(x_0)=(y_{1}+y_{-1}-2y_0)/a^2$ for the first and second derivatives.
}
of the interpolation polynomial $I_{2N}(x_1)|_{n_2,n_3}$ in the $x_1$ coordinate,
uniquely determined by the series of values $A(n_1+n,n_2,n_3)$ $(n=-N,..,N)$ obtained via the averaging (\ref{average}),
and then, finally, taking the limit $N \rightarrow \infty$\footnote{
Note that the above definition of the spatial lattice derivative looks similar to the SLAC-derivative \cite{Drell}, which 
 solves the fermion doubling problem, but has no well-defined continuum limit (see \cite{J.Smit}).
 The SLAC derivative results from our definition (\ref{1stdisc}) with (\ref{distribution}),
 if the limit $N\to\infty$ is taken before the sum over $m$ is carried out.
The expected absence of the fermion-doubling problem is good news for the generalisation of our approach to include
fermions. The absence of a well-defined continuum limit, on the other hand, does not present a problem in our large-box case,
since -- as will be discussed below --we expect that with decreasing length $a$ first a transition -- at some intermediate scale --
to the small-box scenario by L\"uscher \cite{Luescher, Weisz and Ziemann} using the Coulomb-gauge  will be necessary, 
before a well-defined limit $a\rightarrow 0$ can be taken.
 }.
Note, that the ($N=1$) choice,
$\partial_s A({\mathbf{n}})|_{N=1} =\left(A({\mathbf{n}}+ {\mathbf{e}}_s)
     -A({\mathbf{n}}- {\mathbf{e}}_s)\right)/(2a)$,
which includes only the nearest neighbors ${\mathbf{n}}\pm {\mathbf{e}}_s$,
would lead to the same results as (\ref{1stdisc})
for the soft components of the original field $A({\mathbf{x}})$, varying only slightly over several lattice sites, 
but lead to values falling off faster than (\ref{1stdisc}) 
for higher momentum components  close to $\pi/a$. 
For example, we have\footnote{Noting $\lim_{N\to\infty}\sum_{m=1}^N m^{2k} w_N(m) =\delta_{0 k}$.}
\begin{eqnarray}
\lim_{N\to\infty}\sum_{m=1}^N { w_N(m) \over m}\sin{[m(a k)]}=a k ~.
\label{sin[m(a k)]}
\end{eqnarray}

Applying furthermore the rescaling transformation (again afterwards dropping the primes)
\begin{equation}
A = \frac{g^{-1/3}}{a} A^{\prime}~, \quad\quad
P  = \frac{g^{1/3}}{a^2} P^{\prime}~,
\label{rescaling trafo}
\end{equation}
an expansion of the Hamiltonian in $\lambda=g^{-2/3}$ can be obtained
\begin{equation}
H =\frac{g^{2/3}}{a}\left[{\cal H}_0+\lambda \sum_\alpha {\cal V}^{(\partial)}_\alpha
                             +\lambda^2 \sum_\beta {\cal V}^{(\partial\partial)}_\beta
                             + {\mathcal{O}}(\lambda^3)\right]~,
\end{equation}
with the dimensionless and coupling constant independent terms ${\cal H}_0,{\cal V}^{(\partial)}_\alpha, {\cal V}^{(\partial\partial)}_\beta,...\ $.

The "free" part ${\cal H}_0$ 
is the just the sum of the Hamiltonians of $SU(3)$-Yang-Mills quantum mechanics of constant
fields in each box,
\begin{eqnarray}
\label{calH0}
{\cal H}_0 &=&
\sum_{\mathbf{n}}{\cal H}^{QM}_0({\mathbf{n}})~,
\end{eqnarray}
and the interaction parts  
${\cal V}^{(\partial)}_\alpha, {\cal V}^{(\partial\partial)}_\beta ,...$
are relating different boxes. Those contributing up to second order in strong coupling read
\begin{eqnarray}
\label{calVp}
{\cal V}^{(\partial_s)}_\alpha & =&\lim_{N\rightarrow\infty}
          \sum_{m=1}^N \frac{w_N(m)}{2m}
\sum_{\mathbf{n}} \Bigg[
\widetilde{\cal Y}_{\alpha }({\mathbf{n}})
\Big({\cal Y}_{\alpha }({\mathbf{n}}+m{\mathbf{e}}_s)
     -{\cal Y}_{\alpha }({\mathbf{n}}-m {\mathbf{e}}_s)\Big)\ + {\rm h.c.} \Bigg]~,
\\
\label{calVpp}
{\cal V}^{(\partial_s\partial_s)}_\beta &=&\lim_{N\rightarrow\infty}
        \sum_{m,m^\prime=1}^N \frac{w_N(m) w_N(m^\prime)}{4 m m^\prime}
\sum_{{\mathbf{n}}} \Bigg[  \Big({\cal X}_{\beta }({\mathbf{n}}+m{\mathbf{e}}_s)
     - {\cal X}_{\beta }({\mathbf{n}}-m{\mathbf{e}}_s)
    \Big)
\nonumber\\
&&\quad\quad\quad\quad\quad\quad\quad\quad\quad\quad\quad\quad\quad\quad\quad\quad
\widetilde{\cal X}_{\beta}({\mathbf{n}})
\Big({\cal X}^\prime_{\beta }({\mathbf{n}}+m^\prime{\mathbf{e}}_s)
     - {\cal X}^\prime_{\beta }({\mathbf{n}}-m^\prime{\mathbf{e}}_s)
    \Big)\ + {\rm h.c.} \Bigg]~,
\end{eqnarray}
in terms of the dimensionless and coupling constant independent terms 
${\cal X},\widetilde{\cal X},{\cal X}^\prime,\widetilde{\cal Y},{\cal Y}$,
which are obtained from those in (\ref{partial}) and (\ref{partial2}) putting $a=1,g=1,\delta(0)=1$ .

The expansion of the unconstrained $SU(3)$ Yang-Mills Hamiltonian in the number of spatial derivatives
is therefore equivalent to a strong coupling expansion in $\lambda=g^{-2/3}$, just as for the $SU(2)$ case \cite{pavel2010}.
It is the analogon
of the weak coupling expansion in $g^{2/3}$ for small boxes
by L\"uscher \cite{Luescher} for $SU(2)$ Yang-Mills theory and by Weisz and Ziemann\cite{Weisz and Ziemann} 
for $SU(3)$ Yang-Mills theory,
and supplies an useful alternative to strong coupling expansions based on the
Wilson-loop gauge invariant variables, proposed by Kogut, Sinclair, and Susskind
\cite{Kogut} for a 3-dimensional spatial lattice in the Hamiltonian approach,
yielding an expansion in $1/g^4$, and by M\"unster \cite{Muenster} for a
4-dimensional space-time lattice.

\section{The spectrum of $SU(3)$ Yang-Mills quantum mechanics}

The low energy spectrum $\epsilon^{(S)PC}_i ({\mathbf{n}}) $ and eigenstates $|\Psi_{i,M}^{(S)PC}\rangle_{\mathbf{n}}$ 
of ${\cal H}_0^{QM}$  at each site $x_{\mathbf{n}}$ appearing in (\ref{calH0}), are the solutions of the Schr\"odinger
eigenvalue problem of $SU(3)$ Yang-Mills quantum mechanics of spatially constant fields 
\begin{equation}
H_0 (A,P) |\Psi_{i,M}^{(S)PC}\rangle =
\epsilon^{(S)PC}_i \left(\frac{g^{2/3}}{a}\right) |\Psi_{i,M}^{(S)PC}\rangle~,
\end{equation}
with the Hamiltonian (\ref{H_0}), after having applied the rescaling trafo (\ref{rescaling trafo}), and the measure
\begin{equation}
\langle\Psi_1|{\cal O} |\Psi_2\rangle=
\int d\mu_X\int d\mu_Y\int d\mu_Z  \ \Psi^{\dagger}_1 O\ \Psi_2~,
\label{qm-measure3}
\end{equation} 
with the $d\mu_X, d\mu_Y,d\mu_Z$ given by (\ref{qm-measure2}).
The solutions, characterised by the quantum numbers of spin $S,M$, parity $P$, and charge conjugation $C$, 
can in principle be obtained with arbitrary high accuracy. This  is discussed in detail, in particular also the explicit expression 
for the spin-operator in reduced A-space, in \cite{pavel2021}. 
For completeness, I give a summary of some of the main results here.

\subsection{The corresponding harmonic oscillator problem}
Replacing in $ H_0 (A,P)$ of (\ref{H_0}) the magnetic potential by the separable harmonic oscillator potential
 with free parameter $\omega>0$
\begin{eqnarray}
 {1\over 2}   \left(B^{\rm hom}_{ai}(A)\right)^2 
\quad\longrightarrow \quad  {1\over 2}\, \omega^2 \left(A_{ai}\right)^2=
 {1\over 2}\, \omega^2\left(r_X^2+( r_Y^2+r_{1Y}^2+r_{2Y}^2+r_{3Y}^2)+ \sum_{a=1}^8Z_a^2\right)
\end{eqnarray}
the corresponding harmonic oscillator problem (with the same measure (\ref{qm-measure3}))
\begin{equation}
H_{h.o.} (A,P) |\Phi_{i,M}^{(S)PC}\rangle =
\epsilon^{(S)PC}_{h.o.}\left(\frac{g^{2/3}}{a}\right) |\Phi_{i,M}^{(S)PC}\rangle~,
\label{H_0ho}
\end{equation}
becomes trigonal in the space of the monomial functionals
\begin{equation}
 M[\omega s_{[2]},\omega^{3/2} s_{[3]},\omega^{2} b_{[4]},\omega^{5/2} b_{[5]},
\omega^{3/2} a_{[3]},\omega^{2} a_{[4]},\omega^{5/2} a_{[5]},\omega^{3} a_{[6]}]
\, \exp[-\omega\left(s_{11}+s_{22}+s_{33}\right)/2]~,
\label{monomials}
\end{equation}
where the $M$ are monomials in the 45 components of eight elementary $SU(3)$-invariant spatial tensors in reduced A-space
shown in Table 1. For example, $M=1$ or  $M=s_{11}$ or $M=s_{22}^2 s_{123} b_{23}$ .

\begin{table}
$\begin{array}{|c|c|c|c|}  
\hline
{\rm  sym.\ 2\, tensor}&\!\! [2]\!\!& s_{[2]ij}^{++}[A]:=A_{a i}A_{a j}~,\quad (i \le j) &0^{++},2^{++}         
 \\  \hline
{\rm  sym.\ 3\, tensor}&\!\![3]\!\!&s_{[3]ijk}^{--}[A]:=d_{abc}\, A_{a i}A_{b j}A_{c k}~,  \quad (i \le j \le k) &1^{--},3^{--} 
 \\  \hline
{\rm  sym.\ 2\, tensor}&\!\![4]\!\!& b_{[4]ij}^{++}[A]:=B^{\rm hom}_{a i}B^{\rm hom}_{a j}~,\quad (i \le j) 
  ~,\quad\quad   B^{\rm hom}_{a i}:= (1/2) \epsilon_{ijk}\, f_{abc}\, A_{b j}A_{c k} &0^{++},2^{++}   
\\  \hline
{\rm vector}&\!\![5]\!\!& b_{[5]i}^{--}[A]:=d_{abc}\, B^{\rm hom}_{a i}B^{\rm hom}_{b i}A_{c i}
+{1\over 4}\left(2s_{jk}s_{123}-s_{jj}s_{ikk}-s_{kk}s_{ijj}\right),  (\! i\!  \neq\! j\!\! \neq\! k\!) \!\!  &1^{--}    
 \\  \hline
{\rm axial\ scalar}&\!\![3]\!\!&a_{[3]}^{-+}[A]:=f_{abc}\, A_{a 1}A_{b 2}A_{c 3}
=B^{\rm hom}_{a 1}A_{a 1}=B^{\rm hom}_{a 2}A_{a 2}=B^{\rm hom}_{a 3}A_{a 3} &0^{-+}      
 \\  \hline
{\rm axial\ vector}&\!\![4]\!\!& a_{[4]i}^{+-}[A]:=d_{abc}\, B^{\rm hom}_{a i}A_{b i} A_{c i}~,\quad (i=1,2,3) &1^{+-}        
 \\  \hline
{\rm   sym.\,axial\ 2\, tens.}\!\!&\!\![5]\!\!& a_{[5]ij}^{-+}[A]:=d_{abc}\, B^{\rm hom}_{a k} A_{b k} (d_{cde}\, A_{d i}A_{e j})~,
\quad (i \le j \wedge k\neq i,j)    &0^{-+},2^{-+}
 \\  \hline  
 {\rm  sym.\, axial\  3\,  tens.}\!\!&\!\! [6]\!\! &a_{[6]ijk}^{+-}[A]:=
d_{abc}B^{\rm hom}_{a i}B^{\rm hom}_{b j}B^{\rm hom}_{c k}~, \quad (i \le j \le k) &\! 1^{+-}\! ,3^{+-}\!\! 
 \\  \hline
\end{array}$
\caption{Definition of the complete set of eight elementary $SU(3)$-invariant spatial tensors on gauge-reduced A-space. 
The indices $a,b,c$ are summed over, but the spatial indices $i,j,k$ are not, in all lines of the table. 
Note that for the case $i=j$ in the last line one can choose any of the two $ k\neq i,j $, 
both give the same $a_{[5]}$. The second column shows the degree $[n]$ of the tensor (as a polynomial in A).
The last column shows the spin components into which the tensor can be decomposed.}
\end{table}

My result is in accordance with a theorem proven by Dittner \cite{Dittner}, that the 
primitive $SU(3)$-invariant tensors  in 
original constrained V-space have maximal rank 6 and that their number is  35. 
The eight irreducible symmetric spatial tensors in Table 1 are indeed maximally of rank 6.
There is, however a considerable conceptual simplification in reduced $A$ space in comparison to the constrained $V$ space.
For example, in constrained V-space, the $SU(3)$-invariant
\begin{eqnarray}
d_{abc} C_{12}^{a}[V] C_{12}^{b}[V] C_{12}^{c}[V] \quad\quad {\rm with}  \quad \quad                    
 C_{12}^{a}[V]:= d_{abc}\, V^b_{1}\, V^c_{2}~,
\end{eqnarray}
discussed in \cite{Dittner}, is independent of the eight invariants $s_{11}[V]$, $s_{12}[V]$, $s_{12}[V]$, 
$s_{111}[V]$, $s_{112}[V]$, 
$s_{122}[V]$, $s_{222}[V]$, $b_{33}[V]$, using the functionals
 in Table 1 in terms of unreduced $V$ instead of reduced $A$
\footnote{
According to well known identites, $b_{33}[V]\equiv (f_{abc}V_1^{b}V_2^{c}) (f_{ade}V_1^{d}V_2^{e})$ 
according to Table 1 is
related to the invariant $C_{12}^{a}[V]  C_{12}^{a}[V]$ via  $b_{33}[V]\equiv 3\, C_{12}^{a}[V]  C_{12}^{a}[V]
-s_{12}[V]  s_{12}[V]$ }
, in the sense, that it cannot be represented as a sum of products of them. 
It is, however, not primitive because it is related to them via outer products.

In reduced A-space, however, where outer products of invariant tensors are absent, the corresponding polynomial 
is indeed reducible
\begin{eqnarray}
d_{abc} C_{12}^{a}[A] C_{12}^{b}[A] C_{12}^{c}[A] &=&
 {1\over 18} s_{12}^3[A] - {1\over 6} s_{12}[A]\,  s_{11}[A]\,  s_{22}[A]  - {1\over 12} s_{111}[A]\,  s_{222}[A] 
\nonumber\\
&&\quad\quad\quad\quad
+ {3\over 4}  s_{112}[A]\,  s_{122}[A] + 
   {1\over 6} s_{12}[A]\,  b_{33}[A]~.
\end{eqnarray}

Organising the monomial functionals (\ref{monomials}) according to  the degree $n$ (as a polynomial in the $A$) and 
the conserved quantum numbers S,M,P,C
and applying a Gram-Schmidt orthogonalisation with respect to the measure (\ref{qm-measure}), 
we obtain all exact solutions
\begin{equation}
\Phi_{[n]\,i,M}^{(S)PC}[A]= P_{[n]\,i,M}^{(S)PC}[\omega s_{[2]},\omega^{3/2} s_{[3]},\omega^{2} b_{[4]},
\omega^{5/2} b_{[5]},
\omega^{3/2} a_{[3]},\omega^{2} a_{[4]},\omega^{5/2} a_{[5]},\omega^{3} a_{[6]}]\, \exp[-\omega\left(A_{ai}\right)^2/2]~,
\end{equation}
of the corresponding harmonic oscillator problem (\ref{H_0ho}) with energies
\begin{eqnarray}
\epsilon^{(S)PC}_{h.o.} &=& \left(12+n\right)\omega~,
\end{eqnarray}
where $n$ is the degree of $P_{[n]}$ as a polynom in the $A$.

\noindent
 For the lowest $0^{++}$ eigenstates e.g., we find
\begin{eqnarray}
\epsilon^{(0)++}_{h.o.} = 12\,\omega~:
\quad 
P_{[0]}^{(0)++}\!\!\! &\propto&\!\!\! 1~,
\nonumber\\
\epsilon^{(0)++}_{h.o.} = 14\,\omega~:
\quad 
P_{[2]}^{(0)++}\!\!\! &\propto&\!\!\!  -2\sqrt{3}+{1\over 2}\,\omega\, s^{(0)++}_{[2]}~,
\nonumber\\
\epsilon^{(0)++}_{h.o.} = 16\,\omega~:
\quad 
P_{[4]1}^{(0)++}\!\!\! &\propto&\!\!\!  \sqrt{78} -\sqrt{{13\over 2}}\,\omega\, s^{(0)++}_{[2]}
+{1\over 2}\sqrt{{3\over 26}}\,\omega^2  \left( s^{(0)++}_{[2]}\right)^2 ~,
\nonumber\\
\quad 
P_{[4]2}^{(0)++}\!\!\! &\propto&\!\!\! -{1\over 2 \sqrt{273}}\omega^2\left( s^{(0)++}_{[2]}\right)^2
 +{1\over 2}\sqrt{{13\over 105}}\omega^2\langle s^{(2)}_{[2]} s^{(2)}_{[2]}\rangle^{(0)++}~,
\nonumber\\ 
P_{[4]3}^{(0)++}\!\!\! &\propto&\!\!\!
-{1\over 3}\sqrt{{2\over 35}}\omega^2\left( s^{(0)++}_{[2]}\right)^2
+{1\over 3 \sqrt{14}} \omega^2\langle s^{(2)}_{[2]} s^{(2)}_{[2]}\rangle^{(0)++}\!\!\!
+{1\over 9}\sqrt{{14\over 5}}\,\omega^2\, b^{(0)++}_{[4]}~,
\end{eqnarray}
(overall prop. const. $=2\sqrt{2}\,\omega^6/\pi^{5/2}$) and so on, in terms of the spin-0 component
\begin{eqnarray}
s^{(0)++}_{[2]}= {1\over \sqrt{3}}\left(s_{11}+s_{22}+s_{33}\right)
\end{eqnarray}
the five spin-2  components 
\begin{eqnarray}
s^{(2)++}_{[2]\, 0} = {1 \over\sqrt{6} }\left(s_{11}+s_{22}-2\,s_{33}\right)
\quad\quad
s^{(2)++}_{[2]\, \pm 1} =\pm\, s_{13}+ i\, s_{23}
\quad\quad
s^{(2)++}_{[2]\, \pm 2} =- {1\over 2}\left(s_{11}-s_{22}\right)\mp i  s_{12}
\end{eqnarray}
of the 2-tensor $s^{++}_{[2]}$ and similarly for the  2-tensor $b^{++}_{[4]}$.
Much more details are presented in \cite{pavel2021}.

\subsection{Low-energy eigensystem of $SU(3)$ Yang-Mills quantum mechanics}

Consider the  basis of energy eigenstates of the correponding unconstrained harmonic oscillator Schr\"odinger equation 
orthonormal with respect to 
the Yang-Mills measure (\ref{qm-measure3})
\begin{eqnarray}
H_{\rm h.o.}\Phi_n[A,\omega]\equiv \left[T_{\rm kin}+{1\over 2}\omega^2 A_{ai}^2\right]\Phi_n[A,\omega]
=\epsilon^{\rm h.o.}_n \Phi_n[A,\omega]~.
\end{eqnarray}
Then the matrix elements of the unconstrained Yang-Mills Hamiltonian are given as
\begin{eqnarray}
 \!\!\! \!\!\! \!\!\!
{\cal M}_{mn} \!\!\!&:=&\!\!\! \Phi^\dagger_m[A,\omega]\left(T_{\rm kin}+{1\over 2}  B_{ai}^2[A]\right)\Phi_n[A,\omega]
=\delta_{nm}\epsilon^{\rm h.o.}_n
+{1\over 2}\Phi^\dagger_m[A,\omega]\left( B_{ai}^2[A] -\omega^2 A_{ai}^2\right)\Phi_n[A,\omega],
\end{eqnarray}
since the kinetic terms $T_{\rm kin}$ are the same for the Yang-Mills and the corresponding harmonic oscillator problem.
We treat $\omega$ as a variational parameter, which in each symmetry sector can be choose to minimize the lowest
eigenvalue of the matrix ${\cal M}$.
 The results for the energy eigenvalues and eigenstates obtained in the recent work \cite{pavel2021} considerably improve those
obtained by Weisz and Ziemann \cite{Weisz and Ziemann} using the constrained functional approach working 
in unreduced $V$-space .

The spectrum obtained is purely discrete in accordance with the proof by Simon \cite{Simon} and the value 
\begin{equation}
\epsilon_{0}^{++}=12.5868~.
\end{equation}
for the groundstate energy is found. The energies (relative to $\epsilon_{0}$)
\begin{equation}
\mu_{i}^{(S)CP}:=\epsilon_{i}^{(S)CP}-\epsilon_{0}^{++}~,
\end{equation}
of the lowest states are summarized in Table 2. In order to obtain from these "bare" masses
the corresponding physical glueball spectrum, which can be compared e.g. with the lattice results  \cite{Morningstar}\cite{Chen},
a proper renormalisation has to be carried out, in order to remove the dependence on the box-length $a$.
How this could be accomplished, will be discussed in the next Section.

\begin{table}
$\begin{array}{|c||c|c|c|c|c||c|c||c||c|c|c|c|c|}
\mu_{i}^{(S)PC} & 0^{++} & 2^{++}  & 4^{++}  & 6^{++} & 3^{++} & 0^{-+} & 2^{-+}& 1^{+-}& 1^{--} 
& 3^{--} & 5^{--} & 2^{--} & 4^{--}   \\  \hline\hline
i=1 & 2.76 & 2.21 & 4.44 & 6.7 & 6.9 & 5.16 & 7.25 & 6.18 & 3.94 & 3.44 & 5.5 & 5.8 & 5.7 \\ \hline
i=2 & 4.5 & 4.5 & 6.9 & 6.9 & 8.7 & 7.4 & 8.2 & 8.7 & 5.9 & 5.9 & 8  & 7.4 & 8.1   \\ \hline
\end{array}$
\caption{Results for lowest excitation energies $\mu_{i}^{(S)PC}$, the "bare" glueball masses, calculated from the results of \cite{pavel2021}
using orthogonal polynomials up to 10th and 11th order. 
The numerical errors are of the order of the last digit in the numbers given.}
\end{table}

\section{Perturbation theory in $\lambda=g^{-2/3}$}

\subsection{ Free many-glueball states}

The eigenstates of the free Hamiltonian
\begin{eqnarray}
H_0=\frac{g^{2/3}}{a}\sum_{\mathbf{n}}{\cal H}^{QM}_0({\mathbf{n}})
 \nonumber
\end{eqnarray}
are free many-glueball states, i.e. the tensor product of the glueball-eigenstates of completely decoupled granulas.
In particular, the free glueball-vacuum is given as 
\begin{eqnarray}
 |0\rangle \equiv  \bigotimes_{\mathbf{n}} |\Phi_0\rangle_{\mathbf{n}} \
 \rightarrow\  E_{\rm vac}^{\rm free}={\cal N} \epsilon_0\frac{g^{2/3}}{a}~,
 \nonumber
\end{eqnarray}
(${\cal N}$ total number of granulas) with all granulas in the lowest state of energy $\epsilon_0$.
Furthermore, the free one-glueball states, which can be chosen to be e.g. momentum eigenstates, are
\begin{eqnarray}
 |S,M,P,C,i,{\mathbf{k}}\rangle &\equiv & {1\over \sqrt{{\cal N}}} \sum_{\mathbf{n}} e^{i a {\mathbf{k}}.{\mathbf{n}}}
 \left[ |\Phi_{i,M}^{(S)PC}\rangle_{\mathbf{n}} \bigotimes_{{\mathbf{m}}\neq {\mathbf{n}}}
 |\Phi_0\rangle_{\mathbf{m}}\right]\
\rightarrow \
 E^{(S)PC}_{i\ \rm free}(k)
=\mu_i^{(S)PC}\frac{g^{2/3}}{a} + E_{\rm vac}^{\rm free}~,
\nonumber
\end{eqnarray}
the free two-glueball states,
\begin{eqnarray}
|(S_1,M_1,P_1,C_1,i_1,{\mathbf{n}}_1),(S_2,M_2,P_2,C_2,i_2,{\mathbf{n}}_2)\rangle &\equiv &
 |\Phi_{i_1,M_1}^{(S_1)P_1 C_1}\rangle_{{\mathbf{n}}_1}
 \otimes|\Phi_{i_2,M_2}^{(S_2) P_2 C_2}\rangle_{{\mathbf{n}}_2}
\left[ \bigotimes_{{\mathbf{m}}\neq {\mathbf{n}}_1,{\mathbf{n}}_2}
 |\Phi_0\rangle_{\mathbf{m}}\right]\nonumber\\
\rightarrow \
 E_{i_1,i_2\ \rm free}^{(S_1,S_2)\  P_1P_2\ C_1 C_2}
&=&(\mu_{i_1}^{(S_1)P_1 C_1}+\mu_{i_2}^{(S_2)P_2 C_2})\frac{g^{2/3}}{a}
+ E_{\rm vac}^{\rm free}~,\nonumber
\end{eqnarray}
and so on.
Matrix elements between these free many-glueball states are calculated using the measure (\ref{qm-measure})
with the product over space $ {\mathbf{x}}$ replaced by the product over $ {\mathbf{n}}$.

\subsection{Interacting glueball vacuum}

The energy of the interacting glueball vacuum up to $\lambda^2$
\begin{eqnarray}
\!\!\!\!\!\!\!\!\!\!\!\!\!\!\!\!\!\!\!\!\!\!\!\!\!
E_{\rm vac}\!\!\!\!\!&=&\!\!\!\!\! {\cal N}\frac{g^{2/3}}{a}
\Bigg[\epsilon_0+
\lambda^2
\sum_{s=1}^3 \sum_\beta\langle 0 | {\cal V}^{(\partial_s\partial_s)}_\beta|0\rangle
-\lambda^2\sum_{s=1}^3\sum_{\alpha,\alpha^\prime}\sum_{\  i_1, i_2}
\frac{
\langle 0 | {\cal V}^{(\partial_s)}_{\alpha}|i_1, i_2\rangle\langle i_1, i_2|
{\cal V}^{ (\partial_s)}_{\alpha^\prime}|0 \rangle}
{\mu_{i_1}+\mu_{i_2}}
 +{\cal O}(\lambda^3)
\Bigg]
\nonumber\\
&\equiv&
 {\cal N}\frac{g^{2/3}}{a}\Bigg[\epsilon_0+\lambda^2\sum_{s=1}^3 c_{0\,[s]}+{\cal O}(\lambda^3)\Bigg]
\equiv {\cal N}\frac{g^{2/3}}{a}\Bigg[\epsilon_0+\lambda^2c_0+{\cal O}(\lambda^3)\Bigg]~,
\end{eqnarray}
is obtained using first and second order perturbation theory. 
To the simplify the notation we have used  $| i \rangle\equiv |S,M,P,C, i, n\rangle$ and  $\mu_i\equiv \mu_i^{(S)PC}$
for the intermediate states.

For the first order contribution one obtains
\begin{eqnarray}
c^{\ {\rm 1st\ ord.}}_{0\,[s]} 
=\frac{\pi^2}{3} \sum_\beta \langle 0 | \widetilde{\cal X}^{(s)}_\beta|0\rangle
\Big(\langle 0 | {\cal X}^{(s)}_\beta {\cal X}^{\prime (s)}_\beta|0\rangle
-\langle 0 | {\cal X}^{(s)}_\beta |0\rangle \langle 0 | {\cal X}^{\prime (s)}_\beta|0\rangle\Big)~,
\end{eqnarray}
and for the second order contribution
\begin{eqnarray}
c^{\ {\rm 2nd\ ord.}}_{0\,[s]} 
&=& -\frac{\pi^2}{3} \sum_{\alpha,\alpha^\prime}\sum_{\  i_1\neq i_2}
\frac{
\langle 0 | \widetilde{\cal Y}^{(s)}_{\alpha}|i_1\rangle\langle 0 | {\cal Y}^{(s)}_{\alpha}|i_2\rangle\Big(\langle  i_2|
 \widetilde{\cal Y}^{ (s)}_{\alpha^\prime}|0 \rangle \langle  i_1| {\cal Y}^{ (s)}_{\alpha^\prime}|0 \rangle-(i_1\leftrightarrow i_2)\Big)}
{\mu_{i_1}+\mu_{i_2}}~.
\end{eqnarray}
Due to rotational invariance we should find $ c_{0\,[s=1]}=c_{0\,[s=2]}=c_{0\,[s=3]}$ or
\begin{eqnarray}
&& c^{\ {\rm 1st\ ord.}}_{0\,[s=1]}+c^{\ {\rm 2nd\ ord.}}_{0\,[s=1]}=
c^{\ {\rm 1st\ ord.}}_{0\,[s=2]}+c^{\ {\rm 2nd\ ord.}}_{0\,[s=2]}=
c^{\ {\rm 1st\ ord.}}_{0\,[s=3]}+c^{\ {\rm 2nd\ ord.}}_{0\,[s=3]}~.
\end{eqnarray}

\subsection{Interacting glueballs}

Including interactions $V^{\partial\partial}$ and $V^{\partial}$  using 1st and 2nd order perturbation theory,
we can obtain the following energy of the interacting  lowest lying $J^{PC}$ glueballs.
Most glueball excitations are unstable at tree-level, except for the lowest $\mu_1^{(S)PC}$,
which are below threshold  for decay into two spin-2 glueballs.
For example, the energy of the interacting  lowest lying $0^{++}$ glueball up to $\lambda^2$,
(writing $| 1 k \rangle\equiv |0,0,+,+, 1, k\rangle$)
\begin{eqnarray}
E_1^{(0)++}(k)-E_{\rm vac}\!\!\!\! &=&\!\!\!\! \frac{g^{2/3}}{a}
\Bigg\{\epsilon_1^{(0)++} +
\lambda^2\sum_{s=1}^3\sum_\beta
\langle  1 k | {\cal V}^{(\partial_s\partial_s)}_\beta| 1 k\rangle\!
\nonumber\\
&&\quad
-\lambda^2\sum_{s=1}^3\sum_{\alpha,\alpha^\prime}\sum_{i}\!^\prime
\Bigg[\frac{
\langle  1 k | {\cal V}^{(\partial_s)}_{\alpha}|i\rangle\langle i|
{\cal V}^{ (\partial_s)}_{\alpha^\prime}| 1 k \rangle}
{\mu_{i}-\mu_{1}^{(0)++}}+
\frac{
\langle  1 k | {\cal V}^{(\partial_s)}_{\alpha}|1k,i, 1k\rangle\langle 1k,i, 1k|
{\cal V}^{ (\partial_s)}_{\alpha^\prime}| 1 k \rangle}
{\mu_{i}+\mu_{1}^{(0)++}}
\Bigg]
\nonumber\\
&&\quad
-\lambda^2\sum_{s=1}^3\sum_{\alpha,\alpha^\prime}\sum_{ i_1, i_2}\!^\prime
\frac{
\langle  1 k | {\cal V}^{(\partial_s)}_{\alpha}|i_1, i_2\rangle\langle i_1, i_2|
{\cal V}^{ (\partial_s)}_{\alpha^\prime}| 1 k \rangle}
{\mu_{i_1}+\mu_{i_2}-\mu_{1}^{(0)++}}
+{\cal O}(\lambda^3)\Bigg\}
-E_{\rm vac}
\nonumber\\
&&\!\!\!\!\!\!\!\!\!\!\!\!\!\!\!\!\!\!\!\!\!\!\!\!\!\!\!\!\!\!
\equiv 
 \frac{g^{2/3}}{a}\Bigg\{\mu_1^{(0)++}
 + \lambda^2 \sum_{s=1}^3\Bigg[
 \frac{\pi^2}{3} \left(c_{1\,[s]}^{(0)++}+\widetilde{c}_{1\,[s],\infty}^{(0)++}\right)+ 
\left( \widetilde{c}_{1\,[s]}^{(0)++}
- \widetilde{c}_{1\,[s],\infty}^{(0)++}\lim_{N\to\infty} I[N, (a k_s)^2]\,\right) (a k_s)^2 
\Bigg]\!\!
\nonumber\\
&&
+{\cal O}(\lambda^3)\Bigg\}~,
\label{spin0sp}
\end{eqnarray}
using (\ref{sin[m(a k)]})
and the function
\begin{eqnarray}
I[N, x^2] &=&\sum_{m=1}^N {w^2_N(m)\over 4 m^2}\ {1-\cos{[m x]}\over x^2}
=I_{0}[N]-I_{2}[N]\, x^2+I_{4}[N]\, x^4-+....~,
\label{I[N, x^2]}
\end{eqnarray}
with the infinite series of divergent weighted square averages (n=0,1,2,..)
\begin{eqnarray}
 I_{2n}[N]={1\over 4(2n+2)!}\sum_{m=1}^N m^{2n} w^2_N(m)~,\quad\quad \lim_{N\to\infty}I_{2n}[N]=\infty~.
\end{eqnarray}

\noindent
The $N$-independent coefficients $c^{(0)++}_{1\,[s]}$, $\widetilde{c}^{(0)++}_{1\,[s]}$ and
 $\widetilde{c}^{(0)++}_{1\,[s],\infty}$
are products/sums of quantum mechanical expectation-values and transition elements obtained  in standard 
time-independent perturbation theory
(PT) based on the eigensystem of the free Hamiltonian $H_0$, which can be obtained with high accuracy in Sect. 7 .

For the contributions of first order PT,
one obtains (denoting the quantum mechanical  state $|1\rangle \equiv |\Phi_{1}^{(0)++}\rangle$)
\begin{eqnarray}
c^{(0)++}_{1\,[s]\ \rm 1st\ ord.}  
&=&\!\!
 \sum_\beta \Bigg\{\langle 0 | \widetilde{\cal X}^{(s)}_\beta|0\rangle\Bigg[
\Big(\langle 1 | {\cal X}^{(s)}_\beta {\cal X}^{\prime (s)}_\beta|1\rangle
-\langle 0 | {\cal X}^{(s)}_\beta {\cal X}^{\prime (s)}_\beta|0\rangle\Big)
 -\Bigg(\langle 1 | {\cal X}^{(s)}_\beta|0\rangle
\langle 0 | {\cal X}^{\prime (s)}_\beta|1\rangle +( {\cal X} \leftrightarrow  {\cal X}^{\prime })\Bigg)
\nonumber\\
&&\quad\quad\quad\quad\quad\quad\quad\quad
 -\Bigg(\Big(\langle 1| {\cal X}^{(s)}_\beta|1\rangle-\langle 0 | {\cal X}^{(s)}_\beta|0\rangle\Big)
\langle 0 | {\cal X}^{\prime (s)}_\beta|0\rangle +( {\cal X} \leftrightarrow  {\cal X}^{\prime })\Bigg)
\Bigg]
\nonumber\\
&&\quad\quad\quad\quad
+\Big(\langle 1 | \widetilde{\cal X}^{(s)}_\beta|1\rangle-\langle 0 | \widetilde{\cal X}^{(s)}_\beta|0\rangle\Big)
\Bigg[\langle 0 | {\cal X}^{(s)}_\beta {\cal X}^{\prime (s)}_\beta|0\rangle
-\langle 0 | {\cal X}^{(s)}_\beta |0\rangle \langle 0 | {\cal X}^{\prime (s)}_\beta|0\rangle\Bigg]
\Bigg\}~,
\\
\widetilde{c}^{(0)++}_{1\,[s]\  \rm 1st\ ord.} 
&=&\!\!\!\!
\sum_\beta\langle 0 | \widetilde{\cal X}^{(s)}_\beta|0\rangle
\Bigg[
\langle 1 | {\cal X}^{(s)}_\beta |0\rangle \langle 0 | {\cal X}^{\prime (s)}_\beta|1\rangle
+( {\cal X} \leftrightarrow  {\cal X}^{\prime })
\Bigg]~,
\\
\widetilde{c}^{(0)++}_{1\,[s],\infty\ {\rm 1st\ ord.}}
\!\!\!\!\! 
&=&\!\!\!\!
\sum_\beta\Bigg\{\langle 0 | \widetilde{\cal X}^{(s)}_\beta|1\rangle
\Bigg[\langle 1 | {\cal X}^{(s)}_\beta {\cal X}^{\prime (s)}_\beta|0\rangle
-\langle 1 | {\cal X}^{(s)}_\beta |0\rangle \langle 0 |  {\cal X}^{\prime (s)}_\beta|0\rangle\Bigg]
\nonumber\\
&&\quad\quad\quad\quad\quad\quad\quad\quad\quad\quad\quad
+\langle 1 | \widetilde{\cal X}^{(s)}_\beta|0\rangle
\Bigg[\langle 0 | {\cal X}^{(s)}_\beta {\cal X}^{\prime (s)}_\beta|1\rangle
-\langle 0 | {\cal X}^{(s)}_\beta |0\rangle \langle 0 |  {\cal X}^{\prime (s)}_\beta|1\rangle\Bigg]\Bigg\}~,
\end{eqnarray}
and for the  contributions of second order PT
\begin{eqnarray}
c^{(0)++}_{1\,[s]\  \rm 2nd\ ord.} &=&
 -\sum_{\alpha,\alpha^\prime}
\Bigg\{
\sum_{  i\neq 1}\frac{1}{\mu_{i}}\ 
\Big(\langle 1 | \widetilde{\cal Y}^{(s)}_{\alpha}|i\rangle\langle 0 | {\cal Y}^{(s)}_{\alpha}|1\rangle
-( \widetilde{\cal Y} \leftrightarrow  {\cal Y})\Big)
\Big(\langle  i|
 \widetilde{\cal Y}^{ (s)}_{\alpha^\prime}|1 \rangle \langle  1| {\cal Y}^{ (s)}_{\alpha^\prime}|0 \rangle
-( \widetilde{\cal Y} \leftrightarrow  {\cal Y})\Big)
\nonumber\\
&&\quad
+\sum_{  i\neq 1}\frac{1}{\mu_{i}+\mu_{1}^{(0)++}}\ 
\Big(\langle 0| \widetilde{\cal Y}^{(s)}_{\alpha}|i\rangle\langle 0 | {\cal Y}^{(s)}_{\alpha}|1\rangle
-( \widetilde{\cal Y} \leftrightarrow  {\cal Y})\Big)
\Big(\langle  1|
 \widetilde{\cal Y}^{ (s)}_{\alpha^\prime}|0 \rangle \langle  i| {\cal Y}^{ (s)}_{\alpha^\prime}|0 \rangle
-( \widetilde{\cal Y} \leftrightarrow  {\cal Y})\Big)
\nonumber\\
&&\quad
 +\sum_{ i_1, i_2\neq 1}\frac{1}{\mu_{i_1}+\mu_{i_2}-\mu_{1}^{(0)++}} 
\Bigg[
\Big(\langle 1 | \widetilde{\cal Y}^{(s)}_{\alpha}|i_1\rangle\langle 0 | {\cal Y}^{(s)}_{\alpha}|i_2\rangle
-( \widetilde{\cal Y} \leftrightarrow  {\cal Y})\Big)\times
\nonumber\\
&&\quad\quad\quad\quad\quad\quad\quad\quad\quad\quad\quad\quad\quad
\Big(\langle  i_1|\widetilde{\cal Y}^{ (s)}_{\alpha^\prime}|1 \rangle \langle  i_2| {\cal Y}^{ (s)}_{\alpha^\prime}
|0 \rangle
-( \widetilde{\cal Y} \leftrightarrow  {\cal Y})\Big)+(i_1 \leftrightarrow  i_2)\Bigg]
\Bigg\}~,
\\
\widetilde{c}^{(0)++}_{1\,[s]\ \rm 2nd\ ord.}&=&
 \sum_{\alpha,\alpha^\prime} \Bigg\{\sum_{\  i\neq 1}\frac{1}{\mu_{i}-\mu_{1}^{(0)++}}
\Big(\langle 0 | \widetilde{\cal Y}^{(s)}_{\alpha}|i\rangle\langle 1 | {\cal Y}^{(s)}_{\alpha}|0\rangle
-( \widetilde{\cal Y} \leftrightarrow  {\cal Y})\Big)
\Big(\langle  0|
 \widetilde{\cal Y}^{ (s)}_{\alpha^\prime}|1 \rangle \langle  i| {\cal Y}^{ (s)}_{\alpha^\prime}|0 \rangle
-( \widetilde{\cal Y} \leftrightarrow  {\cal Y})\Big)
\nonumber\\
&&\!\!\!\!\!\!\!\!\!\!\!\!\!\!\!
+\sum_{\  i\neq 1}\frac{1}{\mu_{i}+\mu_{1}^{(0)++}}
\Big(\langle 0 | \widetilde{\cal Y}^{(s)}_{\alpha}|i\rangle\langle 0 | {\cal Y}^{(s)}_{\alpha}|1\rangle
-( \widetilde{\cal Y} \leftrightarrow  {\cal Y})\Big)
\Big(\langle  1|
 \widetilde{\cal Y}^{ (s)}_{\alpha^\prime}|0 \rangle \langle  i| {\cal Y}^{ (s)}_{\alpha^\prime}|0 \rangle
-( \widetilde{\cal Y} \leftrightarrow  {\cal Y})\Big)\Bigg\}~,
\\
\widetilde{c}^{(0)++}_{1\,[s],\infty\ {\rm 2nd\ ord.}}\!\!\!\!\!&=&
\sum_{\alpha,\alpha^\prime}\sum_{\  i_1, i_2\neq 1}\frac{1}{\mu_{i_1}+\mu_{i_2}-\mu_{1}^{(0)++}} 
\Bigg[
\Big(\langle 1 | \widetilde{\cal Y}^{(s)}_{\alpha}|i_1\rangle\langle 0 | {\cal Y}^{(s)}_{\alpha}|i_2\rangle
-( \widetilde{\cal Y} \leftrightarrow  {\cal Y})\Big)\times
\nonumber\\
&&\quad\quad\quad\quad\quad\quad\quad\quad\quad\quad\quad\quad\quad\quad
\Big(\langle  i_2|
 \widetilde{\cal Y}^{ (s)}_{\alpha^\prime}|1 \rangle \langle  i_1| {\cal Y}^{ (s)}_{\alpha^\prime}|0 \rangle
-( \widetilde{\cal Y} \leftrightarrow  {\cal Y})\Big)+(i_1 \leftrightarrow  i_2)
\Bigg]~.
\end{eqnarray}
Note, that, due to rotational invariance, the $s=1,2,3$ contributions  to the coefficients $c^{(0)++}_{1\,[s]}$,
$\widetilde{c}^{(0)++}_{1\,[s]}$ and $\widetilde{c}^{(0)++}_{1\,[s],\infty}$ (adding up first and 
second order perturbation
theory) should give equal contributions,
\begin{eqnarray}
&& \!\!\!\!\!\!\!\!\!\!\!\!\!\!\!\!\!\!\!\!\!
c^{(0)++}_{1\,[s=1]\ {\rm 1st\ ord.}}+c^{(0)++}_{1\,[s=1]\ {\rm 2nd\ ord.}}=
 c^{(0)++}_{1\,[s=2]\ {\rm 1st\ ord.}}+c^{(0)++}_{1\,[s=2]\ {\rm 2nd\ ord.}}=
 c^{(0)++}_{1\,[s=3]\ {\rm 1st\ ord.}}+c^{(0)++}_{1\,[s=3]\ {\rm 2nd\ ord.}}~,
\\
&& \!\!\!\!\!\!\!\!\!\!\!\!\!\!\!\!\!\!\!\!\!
\widetilde{c}^{(0)++}_{1\,[s=1]\ {\rm 1st\ ord.}}+\widetilde{c}^{(0)++}_{1\,[s=1]\ {\rm 2nd\ ord.}}=
\widetilde{c} ^{(0)++}_{1\,[s=2]\ {\rm 1st\ ord.}}+\widetilde{c}^{(0)++}_{1\,[s=2]\ {\rm 2nd\ ord.}}=
\widetilde{c} ^{(0)++}_{1\,[s=3]\ {\rm 1st\ ord.}}+\widetilde{c}^{(0)++}_{1\,[s=3]\ {\rm 2nd\ ord.}}~,
\label{rotinv}
\end{eqnarray}
and an analogous expression for  $\widetilde{c}^{(0)++}_{1\,[s],\infty}$.

Furthermore, in order for the result (\ref{spin0sp}) to be finite in the limit $N\to\infty$, and hence renormalisable,
 the above coefficients $\widetilde{c}^{(0)++}_{1,\infty}$
should vanish, i.e.
\begin{eqnarray}
\widetilde{c}^{(0)++}_{1\,[s],\infty}=\widetilde{c}^{(0)++}_{1\,[s],\infty\ {\rm 1st\ ord.}}
+\widetilde{c}^{(0)++}_{1\,[s],\infty\ {\rm 2nd\ ord.}}\equiv 0 \quad\quad {\rm for\ all}\  s=1,2,3~.
\label{renromcond}
\end{eqnarray}
Denoting in this case, when the conditions of rotational invariance (\ref{rotinv}) and finiteness  (\ref{renromcond}) 
are indeed fulfilled,
\begin{eqnarray}
c^{(0)++}_{1}&:=&\sum_{s=1}^3 c^{(0)++}_{1\,[s]}~,
\\
\widetilde{c}^{(0)++}_{1}&:=&\widetilde{c}^{(0)++}_{1\,[s=1]}=\widetilde{c}^{(0)++}_{1\,[s=2]}
=\widetilde{c}^{(0)++}_{1\,[s=3]}~,
\end{eqnarray}
the expression (\ref{spin0sp}) can be written as
\begin{eqnarray}
E_1^{(0)++}(k)-E_{\rm vac}&=& 
 \frac{g^{2/3}}{a}\Bigg\{\mu_1^{(0)++}
 + \lambda^2\Bigg[
 \frac{\pi^2}{3}c_{1}^{(0)++}+ 
 \widetilde{c}_{1}^{(0)++} (a k)^2 
\Bigg]\!\!
+{\cal O}(\lambda^4)\Bigg\}~.
\label{spin0spfinite}
\end{eqnarray}

Finally, in order for the result (\ref{spin0sp}) to be Lorentz invariant, i.e. to fulfill the energy momentum relation 
for a massive spin-0  particle
$$E=\sqrt{m^2+(ak)^2}= m+(1/(2m))(ak)^2+O((ak)^4) $$
we should have the relation
\begin{eqnarray}
\widetilde{c}^{(0)++}_{1}=\widetilde{c}^{(0)++}_{1\ {\rm 1st\ ord.}}+\widetilde{c}^{(0)++}_{1\ {\rm 2nd\ ord.}}
\equiv 1/(2\mu_{1}^{(0)++})~.
\label{Lorentzcond}
\end{eqnarray}

The check of these conditions, rotational invariance, finiteness and Lorentz invarinace, to order $\lambda^2$, 
 the coefficients $c^{(0)++}_{1\,[s]}$, $\widetilde{c}^{(0)++}_{1\,[s]}$ and $\widetilde{c}^{(0)++}_{1\,[s],\infty}$,
 and hence the quantum  mechanical matrix elements,
have to be calculated explicitly, which is expected to be a rather lengthy, but a very important calculation and 
in my opinion managable task
for future work. Up to now, to the best of my knowledge, nobody can exclude by general arguments the existence 
of the above 
quite naturally arising strong coupling scheme.

\subsection{Glueball-glueball scattering amplitude}

Let us consider the eleastic scattering of two identical lowest lying $0^{++}$ glueball excitations with equal and 
opposite momenta ${\mathbf{k}}^{(2)}=-{\mathbf{k}}^{(1)}$ to the
new momenta  ${\mathbf{k}}^{(2)\prime}=-{\mathbf{k}}^{(1)\prime}$, corresponding to vanishing 
center-of mass momentum 
relative to the lattice,
\begin{eqnarray}
{\mathbf{K}}\equiv\left({\mathbf{k}}^{(1)}+{\mathbf{k}}^{(2)}\right)/2 =0~,
\quad\quad  
{\mathbf{K}^\prime}\equiv\left({\mathbf{k}}^{(1)\prime}+{\mathbf{k}}^{(2)\prime}\right)/2=0~,
\end{eqnarray}
Furthermore, elasticity and hence energy conservation requires 
\begin{eqnarray}
|{\mathbf{k}}^{(1)\prime}|=|{\mathbf{k}}^{(1)}|=:k ~.
\end{eqnarray}
Due to global isotropy, the scattering amplitude defined by
\begin{eqnarray}
\!\!\!\!\!\!\!\!\!\!
{\cal A}({\mathbf{k}}^{(1)\prime}-{\mathbf{k}}^{(1)})\!\!\!\!\! &:=&\!\!\!\!\!
\langle 1~ 1, {\mathbf{K}}^\prime=0,{\mathbf{k}}^{(1)\prime}~ |H|
1~1, {\mathbf{K}}=0,{\mathbf{k}}^{(1)}\rangle
-\langle 1~ 1, {\mathbf{K}}^\prime=0,{\mathbf{k}}^{(1)}~ |H|
1~1, {\mathbf{K}}=0,{\mathbf{k}}^{(1)}\rangle~,
\end{eqnarray}
should depend only on 
\begin{eqnarray}
{\mathbf{q}}:= {\mathbf{k}}^{(1)\prime}-{\mathbf{k}}^{(1)}~.
\end{eqnarray}
Up to order $\lambda^2$, using first and second order perturbation theory, it is found to be
\begin{eqnarray}
{\cal A}({\mathbf{q}})={1\over {\cal N}}\frac{g^{2/3}}{a}
\Bigg\{
\lambda^2
\left[ {1\over 2}
\sum_{s=1}^3 a^2\, q_s ^2\left( \widetilde{d}_{1\,[s]}
+ \widetilde{d}_{1\,[s],\infty}\lim_{N\to\infty}I[N,a^2\, q_s ^2]\,\right) \right]
+{\cal O}(\lambda^4)
\Bigg\}~,
\label{scattering}
\end{eqnarray}
with the same function $I[N,x^2]$ defined in  (\ref{I[N, x^2]}), tending to infinity for $N\to\infty$ and 
with the first order coefficients
\begin{eqnarray}
\widetilde{d}_{1\,[s]}^{{\ \ \rm 1st\ ord.}}&=&
\sum_\beta\langle 0 | \widetilde{\cal X}^{(s)}_\beta|0\rangle
\langle 1 | {\cal X}^{(s)}_\beta |1\rangle \langle 1 | {\cal X}^{\prime (s)}_\beta|1\rangle~,
\\
\widetilde{d}_{1\,[s],\infty}^{\ \ {\rm 1st\ ord.}}&=&
\sum_\beta\langle 1 | \widetilde{\cal X}^{(s)}_\beta|1\rangle
\Bigg[\langle 1 | {\cal X}^{(s)}_\beta {\cal X}^{\prime (s)}_\beta|1\rangle
-\langle 1 | {\cal X}^{(s)}_\beta |1\rangle \langle 0 |  {\cal X}^{\prime (s)}_\beta|0\rangle
-\langle 0 | {\cal X}^{(s)}_\beta |0\rangle \langle 1 |  {\cal X}^{\prime (s)}_\beta|1\rangle
\Bigg]~,
\end{eqnarray}
and the second order coefficients
\begin{eqnarray}
\widetilde{d}_{1\,[s]}^{\ \ {\rm 2nd\ ord.}}&=&0~,
\\
\widetilde{d}_{1\,[s],\infty}^{\ \ {\rm 2nd\ ord.}}&=&
\sum_{\alpha,\alpha^\prime}\sum_{\ i_1> i_2> 1\ }\frac{1}{\mu_{i_1}+\mu_{i_2}-2\mu_{1}^{(0)++}} 
\Big(\langle 1 | \widetilde{\cal Y}^{(s)}_{\alpha}|i_1\rangle\langle 1 | {\cal Y}^{(s)}_{\alpha}|i_2\rangle
-( \widetilde{\cal Y} \leftrightarrow  {\cal Y})\Big)
\nonumber\\
&&\quad\quad\quad\quad\quad\quad\quad\quad\quad\quad\quad\quad\quad\quad\quad\quad
\Big(\langle  i_2|
 \widetilde{\cal Y}^{ (s)}_{\alpha^\prime}|1 \rangle \langle  i_1| {\cal Y}^{ (s)}_{\alpha^\prime}|1 \rangle
-( \widetilde{\cal Y} \leftrightarrow  {\cal Y})\Big)~.
\end{eqnarray}
As for the case of the single interacting glueball, there are constraints from the conditions of isotropy and finiteness 
Firstly, due to rotational invariance, the $s=1,2,3$ contributions  to the coefficients 
$\widetilde{d}_{1\,[s]}$ and $\widetilde{d}_{1\,[s],\infty}$ (adding up first and second order perturbation
theory) should give equal contributions. Secondly,
in order for the result (\ref{scattering}) to be finite in the limit $N\to\infty$, and hence renormalisable,
 the above coefficients $\widetilde{d}_{1\,[s],\infty}$
should vanish, i.e.
\begin{eqnarray}
\widetilde{d}_{1\,[s],\infty}=\widetilde{d}_{1\,[s],\infty}^{\ \ {\rm 1st\ ord.}}
+\widetilde{d}_{1\,[s],\infty}^{\ {\rm 2nd\ ord.}}\equiv 0 \quad\quad {\rm for\ all}\  s=1,2,3~.
\label{renromcond2}
\end{eqnarray}
Hence, in this case, defining 
$$\widetilde{d}_{1}:=\widetilde{d}_{1\,[s=1]}=\widetilde{d}_{1\,[s=2]}=\widetilde{d}_{1\,[s=3]}~,$$
the scattering amplitude becomes
\begin{eqnarray}
{\cal A}({\mathbf{q}})=
{1\over {\cal N}}\frac{g^{2/3}}{a}\Bigg\{\lambda^2
\left[ {1\over 2}
 \widetilde{d}_{1}\,  (a q)^2\right]
+{\cal O}(\lambda^3)\Bigg\}~,
\label{scatteringfinite}
\end{eqnarray}
in lowest order depending only on 
$|{\mathbf{q}}|=:2 k \sin\left(\theta/2\right)$.
Performing the corresponding integrations and using an appropriate overall normalisation, we obtain
an expansion of the physical coupling constant $\lambda_R$ in terms of the original bare $\lambda$
\begin{eqnarray}
\lambda_R=\lambda+c \lambda^2+c^\prime \lambda^3+...~,
\label{lambda_R}
\end{eqnarray}
with some constants $c,c^\prime,...$.

\section{Conclusions}

It has been shown in this work, how the gauge invariant formulation of low energy $SU(2)$ Yang-Mills theory on a
three dimensional spatial lattice, obtained by  replacing integrals by sums
and spatial derivatives by differences, proposed in the earlier work \cite{pavel2010}, can be generalised to $SU(3)$
with a simple, but highly non-trivial, FP-operator,  FP-determinant and inverse FP-operator, which are practically managable.

This has been achieved proposing a new gauge, the "flux-tube gauge", defined in (\ref{gauge cond}), which is shown 
to exist by construction.
In contrast to the $SU(3)$ symmetric gauge, the Faddeev-Popov operator, its determinant and inverse, are rather simple
in the flux-tube gauge, but show a highly non-trivial periodic structure of six Gribov-horizons separating six Weyl-chambers.
Such a Weyl structure in the context of models of the QCD-vacuum has been discussed recently 
e.g. in \cite{Nedelko and Voronin}.

Furthermore, as for the case of $SU(2)$ Yang-Mills theory in the symmetric gauge, the flux-tube gauge allows 
for a systematic and practical 
strong coupling expansion of the $SU(3)$ Hamiltonian in $\lambda\equiv g^{-2/3}$, 
equivalent to an expansion in the number of spatial derivatives.

Constructing the corresponding physical quantum Hamiltonian of $SU(3)$ Yang-Mills theory in the flux-tube gauge
according to the general scheme given by Christ and Lee, a systematic and practical 
expansion of the  $SU(3)$ Hamiltonian in the number of spatial derivatives could be obtained here.
Introducing an infinite spatial lattice with box length $a$, the "free part" is the sum of Hamiltonians of 
Yang-Mills quantum mechanics
of constant fields for each box with a purely discrete spectrum("free glueballs"), 
and the "interaction terms" contain higher and higher number
of spatial derivatives connecting different boxes. 
This expansion has been carried out here explicitly and shown to be equivalent to a strong coupling expansion
in $\lambda=g^{-2/3}$ for large box sizes $a$. It is the analogon
to the weak coupling expansion in $g^{2/3}$ by L\"uscher, applicable for small boxes.

The energy eigensystem of the gauge reduced Hamiltonian of $SU(3)$ Yang-Mills mechanics of spatially constant
fields can be calculated in principle with arbitrary high precision using the orthonormal basis of all solutions of the 
corresponding
harmonic oscillator problem,  which turn out to be made of orthogonal polynomials of the 45 components of eight 
irreducible symmetric
spatial tensors. Rather accurate first results for the lowest bare glueball masses have been obtained in recent work \cite{pavel2021} 
and substantially improve those obtained by Weisz and Ziemann using the unreduced approach.

Thus, the  gauge reduced approach using the flux-tube gauge proposed here, is expected to enable one to obtain 
valuable non-perturbative information about low-energy glueball dynamics,
carrying out perturbation theory in $\lambda$.

Finally, I would like to point out, that the flux-tube gauge exists for low energy Hamiltonian formulations of D+1 dimensional 
$SU(3)$ Yang-Mills theories with $D\ge 2$ and  for strong coupling path-integral formulations 
of $D\ge 2$ dimensional Euclidean $SU(3)$ Yang-Mills theories.
In particular, it might be useful for the gauge-invariant investigation of the strong coupling limit of the Hamiltonian 
of $2+1$ dimensional 
$SU(3)$ Yang-Mills theories (and compare e.g. with the continuum Hamiltonian formulation in terms 
of gauge invariant variables \cite{Nair}).

\section*{Acknowledgements}

I would like to thank A. Dorokhov, S. Nedelko, F. Niedermayer, M. Staudacher, and P. Weisz
for interesting discussions. This work was partly financed by the SFB 647 "Raum-Zeit-Materie: 
Analytische und Geometrische Structuren."

\begin{appendix}
\section*{Appendix}
\section{ On the existence of the flux-tube gauge}
In this appendix will be discussed, under which conditions the flux-tube gauge (\ref{gauge cond}) 
\begin{eqnarray}
\chi_a(A)=\left(\Gamma_i\right)_{ab}A_{b i}({\mathbf{x}})=0\quad a=1,...,8~,
\nonumber
\end{eqnarray}
exists, as a generalisation of the proof for the case of the $SU(2)$ symmetric gauge in App.B of \cite{KMPR}.

A gauge $\chi_a(A)=0$ exists, iff
\begin{eqnarray}
\chi_a(V^{\omega})=0 \quad a=1,...,8~,
\end{eqnarray}
for the gauge transformed 
\begin{eqnarray}
V_{a i}^{\omega}({\mathbf{x}}) \lambda_a/2  &=&
U^{\dagger}[\omega({\mathbf{x}})] \left(V_{b i}({\mathbf{x}}) \lambda_b/2+
{i\over g}\partial_i\right) U[\omega({\mathbf{x}})]~,
\end{eqnarray}
has a unique solution for $\omega({\mathbf{x}})$.
The flux-tube gauge therefore exists iff an  arbitrary gauge potential $V_{ai}({\mathbf{x}})$ can be made to fulfill 
$\left(\Gamma_i\right)_{ab}A_{b i}({\mathbf{x}})=0\quad a=1,...,8~,$ by a unique time-independent gauge 
transformation.
We can write the equations which determine the corresponding gauge transformation $\omega({\mathbf{x}})$ as
\begin{eqnarray}
 {1\over 2}\left(\Gamma_i\right)_{ab} 
{\rm Tr}\left[\lambda_b\, U^{\dagger}(\omega({\mathbf{x}})) V_{i}({\mathbf{x}})\,  U(\omega({\mathbf{x}}))\right] &=&
{1\over g}\left(\Gamma_i\right)_{ab}\Sigma_{bi}(\omega({\mathbf{x}}))~,  \quad a=1,...,8~,
\label{trafo-cond}
\end{eqnarray}
with  
\begin{eqnarray}
\Sigma_{ai}(\omega({\mathbf{x}})):=
{1\over 2i} {\rm Tr}\left[\lambda_a U^{\dagger}(\omega({\mathbf{x}}))\partial_i U(\omega({\mathbf{x}}))\right]~.
\end{eqnarray}

\subsection{Solution of the corresponding infinite-coupling problem}

Consider first the homogeneous problem obtained  in the infinite coupling limit $1/g\rightarrow 0$,
\begin{equation}
 \left(\Gamma_i\right)_{ab} {\rm Tr}\left[\lambda_b\, U^{(0)\dagger} V_{i}\,  U^{(0)}\right]=0~,\quad \forall a=1,...,8~.
\label{homog.equ.}
\end{equation}
This is solved iff we can find a unique unitary transformation $U^{(0)}$ such that the $V_i$ can be rotated into new $A_i$, which
satisfy the gauge condition,
\begin{eqnarray}
U^{(0)\dagger} V_{ i}\, U^{(0)}=A_{i}~,\quad\quad
 \left(\Gamma_i\right)_{ab}A_{bi}=0~,\quad \forall a=1,...,8~,
\end{eqnarray}
It is unique, iff the corresponding homogeneous FP operator  $\gamma(A)$ defined in (\ref{homogFP}) is invertible \cite{Christ and Lee},
\begin{eqnarray} 
 \det(\gamma(A))\neq 0 \quad\quad \gamma_{ab}(A)\equiv \left(\Gamma_i\right)_{ad} \, f_{dbc}A_{ci}~.
\end{eqnarray}

For the case of the flux-tube gauge, the homogeneous equation (\ref{homog.equ.}) reads more explicitely 
\begin{eqnarray}
 {\rm Tr}\left[\lambda_a\, U^{(0)\dagger} V_{1}\,  U^{(0)}\right]=0\quad\quad \forall a=1,2,4,5,6,7 \quad \quad 
\wedge  \quad \quad  {\rm Tr}\left[\lambda_a\, U^{(0)\dagger} V_{2}\,  U^{(0)}\right]=0\quad\quad \forall a=5,7~.
\label{trafo-condU0fluxtube}
\end{eqnarray}
This can always be achieved: In a first step, one diagonalises the spatial 1-component of the gauge
field in the fundamental representation, $A_{a1}=0$ for all $a=1,2,4,5,6,7$. In a second step, one uses the remaining
gauge-freedom, generated by $\lambda_3$ and $\lambda_8$, which leave the $a=3,8$ components of all $A_{ai}$ for all $i=1,2,3$ 
unchanged, in order to put  the spatial 2-components $A_{a2}=0$ for $a=5,7$.
Hence (\ref{trafo-condU0fluxtube}) has the solution
\begin{equation}
U^{(0)}(\omega_1,...,\omega_8) 
\equiv U_2^{(0)}(\omega_3^{\prime\prime},\omega_8^{\prime\prime})\, U_1^{(0)}(\omega_1^\prime,...,\omega_8^\prime)
\end{equation}
with
\begin{equation}
U_1^{(0)}(\omega^\prime): =\exp[i\sum_{a}\omega^\prime_a\lambda_a/2]            
 \quad \quad\quad  
U_2^{(0)}(\omega^{\prime\prime}):=\exp[i\omega^{\prime\prime}_3\lambda_3/2]\,\exp[i\omega^{\prime\prime}_8\lambda_8/2]
\end{equation}
such that
\begin{eqnarray}
U_1^{(0)\dagger} V_{ 1} U_1^{(0)}=A_{1}~,
 \quad \quad
U_2^{(0)\dagger}U_1^{(0)\dagger} V_{2}\, U_1^{(0)} U_2^{(0)} =A_{2}   ~.
\end{eqnarray}
using the invariance of $A_{1}$ under $U_2$ transformations
$U_2^{(0)\dagger} A_{1} U_2^{(0)}=A_{1} $.
It is unique, iff the corresponding homogeneous FP determinant  (\ref{det-gamma-hom}) is invertible
\begin{eqnarray} 
\det(\gamma(A))\propto A_{31}^2\left(A_{31}^2-3\,A_{81}^{2}\right)^2 A_{42} A_{62}\neq 0~.
\end{eqnarray}

\subsection{Solution of  Equ.(\ref{trafo-cond}) using a strong-coupling expansion}

In order to discuss the solulability of the full  Equ.(\ref{trafo-cond}) the following Theorem will be proven:
\newline
\newline
\noindent
{\it Theorem}: Equ.(\ref{trafo-cond}) has a unique solution in the form of a $1/g$ expansion
\begin{eqnarray}
 U(\omega({\mathbf{x}}))= U^{(0)}({\mathbf{x}})\left[1+\sum_{n=1}^\infty \left( {1\over g}\right)^n X^{(n)}({\mathbf{x}})\right]~,
\label{Uexpansion}
\end{eqnarray}
iff the corresponding homogeneous FP operator is invertible at each ${\mathbf{x}}$, i.e. $\det(\gamma(A({\mathbf{x}})))\neq 0$.
\newline
\newline
\noindent
{\it Proof}: Equating equal powers of $1/g$ in the unitarity condition $U^{\dagger}\, U=U\, U^{\dagger}=1$
lead to the condition of unitarity of $ U^{(0)}$
\begin{eqnarray}
 U^{(0)\dagger}\, U^{(0)}&=&U^{(0)}\, U^{(0)\dagger}=1~, 
\label{U0unitarity}
\end{eqnarray}
as well as the conditions
\begin{eqnarray}
X^{(1)}+X^{(1)\dagger}
&=&
0 ~, 
\nonumber\\
X^{(2)}+X^{(2)\dagger}+X^{(1)}X^{(1)\dagger}
&=&
0 ~, 
\nonumber\\
...   ...  & & ...   
\nonumber\\
X^{(n)}+X^{(n)\dagger}+\sum_{i+j=n}X^{(i)}X^{(j)\dagger}
&=&
0 ~,
\nonumber\\
...   ...  & & ...   
\label{UunitarityX}
\end{eqnarray}
for the unknown $X^{(n)}$.
Furthermore, inserting (\ref{Uexpansion}) and equating equal powers of  $1/g$, one finds
that the leading order unitary matrix $U^{(0)}$ should satisfy
\begin{equation}
 \left(\Gamma_i\right)_{ab} {\rm Tr}
\left[\lambda_b\, U^{(0)\dagger}({\mathbf{x}})\, V_{i}({\mathbf{x}})\,  U^{(0)}({\mathbf{x}})\right]=0~,\quad \forall a=1,...,8~,
\label{trafo-condU0}
\end{equation}
and the $X^{(n)}$ fulfill the infinite set of equations at each ${\mathbf{x}}$ 
\begin{eqnarray}
 {1\over 2} \left(\Gamma_i\right)_{ab} {\rm Tr}\left[\lambda_b\left( X^{(1)\dagger} U^{(0)\dagger} V_{i}\,  U^{(0)}
+ U^{(0)\dagger} V_{i}\,  U^{(0)}X^{(1)}\right)\right]
\!\!\!\!\!\!\!\!\!&=&\!\!\!\!\!\!\!\!\! 
\left(\Gamma_i\right)_{ab}\Sigma_{bi}^{(0)}~,
\nonumber\\
 {1\over 2} \left(\Gamma_i\right)_{ab} {\rm Tr}\left[\lambda_b\left( X^{(2)\dagger} U^{(0)\dagger} V_{i}\,  U^{(0)}
+ U^{(0)\dagger} V_{i}\,  U^{(0)}X^{(2)}
+X^{(1)\dagger}U^{(0)\dagger} V_{i}\,  U^{(0)}X^{(1)}\right)\right]
\!\!\!\!\!\!\!\!\!&=&\!\!\!\!\!\!\!\!\! 
\left(\Gamma_i\right)_{ab}\Sigma_{bi}^{(1)}~,
\nonumber\\
...   ...  &\quad\quad\quad & ...   ... 
\nonumber\\
 {1\over 2} \left(\Gamma_i\right)_{ab} {\rm Tr}\left[\lambda_b\left( X^{(n)\dagger} U^{(0)\dagger} V_{i}\,  U^{(0)}
+ U^{(0)\dagger} V_{i}\,  U^{(0)}X^{(n)}
+\sum_{i+j=n}X^{(i)\dagger}U^{(0)\dagger} V_{i}\,  U^{(0)}X^{(j)}\right)\right]
\!\!\!\!\!\!\!\!\!&=&\!\!\!\!\!\!\!\!\! 
\left(\Gamma_i\right)_{ab}\Sigma_{bi}^{(n-1)}~,
\nonumber\\
...   ...  &\quad\quad\quad & ...   ... 
\label{trafo-condX}
\end{eqnarray}
using the $1/g$ expansion
\begin{eqnarray}
\Sigma_{ai}(\omega({\mathbf{x}}))=\sum_{n=0}^\infty \left( {1\over g}\right)^n\!\!\Sigma_{ai}^{(n)}({\mathbf{x}})~.
\label{Sigmaexpansion}
\end{eqnarray}
Note, that the $n$th term in (\ref{Sigmaexpansion}) is given in terms of $U^{(0)}$ and $X^{(1)},..,X^{(n-1)}$.

Hence we find that using the expansion  (\ref{Uexpansion}) of $U$ in $1/g$ the solution of (\ref{trafo-cond}) reduces
to the algebraic problem (\ref{U0unitarity})-(\ref{trafo-condX}).
We now assume that for a given choice of $V_{ai}({\mathbf{x}})$, the first, homogeneous equation (\ref{trafo-condU0}) 
has a unique unitary solution $U^{(0)}$ such that
\begin{eqnarray}
U^{(0)\dagger}({\mathbf{x}}) V_{ i}({\mathbf{x}}) U^{(0)}({\mathbf{x}})=A_{i}({\mathbf{x}})  \quad \wedge  \quad
 \left(\Gamma_i\right)_{ab}A_{bi}({\mathbf{x}})=0
\quad \wedge  \quad
 \det(\gamma(A({\mathbf{x}})))\neq 0 ~.
\end{eqnarray}
Using this solution and the conditions (\ref{UunitarityX}) to express $X^{(n)\dagger}$ in terms of $X^{(n)}$ 
plus terms containing $X^{(1)},...X^{(n-1)}$
one can then rewrite the remaining equations (\ref{trafo-condX})
into the set of equations for $X^{(n)}$
\begin{eqnarray}
 \left(\Gamma_i\right)_{ab} {\rm Tr}\left[\lambda_b\left(- X^{(1)}({\mathbf{x}})  A_{i}({\mathbf{x}})\, 
+ A_{i}({\mathbf{x}})\, X^{(1)}({\mathbf{x}})\right)\right]
\!\!\!\!\!\!\!\!\!&=&\!\!\!\!\!\!\!\!\! 
C_{a}^{(0)}({\mathbf{x}})~,
\nonumber\\
 \left(\Gamma_i\right)_{ab} {\rm Tr}\left[\lambda_b\left(- X^{(2)}({\mathbf{x}})  A_{i}({\mathbf{x}})\, 
+ A_{i}({\mathbf{x}})\, X^{(2)}({\mathbf{x}})\right)\right]
\!\!\!\!\!\!\!\!\!&=&\!\!\!\!\!\!\!\!\! 
C_{a}^{(1)}({\mathbf{x}})~,
\nonumber\\
...   ...  &\quad\quad\quad & ...   ... 
\nonumber\\
 \left(\Gamma_i\right)_{ab} {\rm Tr}\left[\lambda_b\left(- X^{(n)}({\mathbf{x}})  A_{i}({\mathbf{x}})\, 
+ A_{i}({\mathbf{x}})\, X^{(n)}({\mathbf{x}})\right)\right]
\!\!\!\!\!\!\!\!\!&=&\!\!\!\!\!\!\!\!\! 
C_{a}^{(n-1)}({\mathbf{x}})~,
\nonumber\\
...   ...  &\quad\quad\quad & ...   ... 
\label{C-eq}
\end{eqnarray}
where the $n$th order $C^{(n)}$ is given in terms of $U^{(0)}$ and  $X^{(1)},...X^{(n-1)}$.
In terms of $X^{(n)}_a := (1/2){\rm Tr}\left[\lambda_a\, X^{(n)}\right]$, and using  the homogeneous FP operator $\gamma(A)$, 
Equs. (\ref{C-eq}) can be written in the final form
\begin{eqnarray}
i\gamma_{ab}(A({\mathbf{x}}))\, X^{(1)}_b({\mathbf{x}})
\!\!\!\!\!\!\!\!\!&=&\!\!\!\!\!\!\!\!\! 
C_{a}^{(0)}({\mathbf{x}})~,
\nonumber\\
i\gamma_{ab}(A({\mathbf{x}}))\, X^{(2)}_b({\mathbf{x}})
\!\!\!\!\!\!\!\!\!&=&\!\!\!\!\!\!\!\!\! 
C_{a}^{(1)}({\mathbf{x}})~,
\nonumber\\
...   ...  &\quad\quad\quad & ...   ... 
\nonumber\\
i\gamma_{ab}(A({\mathbf{x}}))\, X^{(n)}_b({\mathbf{x}})
\!\!\!\!\!\!\!\!\!&=&\!\!\!\!\!\!\!\!\! 
C_{a}^{(n-1)}({\mathbf{x}})~,
\nonumber\\
...   ...  &\quad\quad\quad & ...   ... 
\label{C-eq-gamma}
\end{eqnarray}
This system of Equs. (\ref{C-eq-gamma}) has a unique solution, solving subsequently for $ X^{(1)}$, $ X^{(2)}$, and so on,
 iff the homogeneous FP operator is invertible, which completes the proof.

\end{appendix}



\begin{thebibliography}{99}
%
\bibitem{Christ and Lee}
N.H. Christ and T.D. Lee, Phys. Rev. D 22 (1980) 939.
%
\bibitem{Luescher}
M. L\"uscher , Nucl. Phys. B219 (1983) 233.
%
\bibitem{Weisz and Ziemann}
P. Weisz and V. Ziemann, Nucl. Phys. B284 (1987) 157.
%
\bibitem{KP1}
A.M. Khvedelidze and H.-P. Pavel,
Phys. Rev. D 59 (1999) 105017.
%
\bibitem{KMPR}
A.M. Khvedelidze, D. M. Mladenov, H.-P. Pavel, and G. R\"opke,
Phys. Rev. D 67 (2003) 105013.
%
\bibitem{pavel2010}
H.-P. Pavel, Phys. Lett. B 685 (2010) 353-364.
%
\bibitem{pavel2007}
H.-P. Pavel,
Phys. Lett. B 648 (2007) 97-106.
%
%
\bibitem{pavel2011}
H.-P. Pavel, 
Phys. Lett. B 700 (2011) 265-276.
%
\bibitem{Creutz}
M. Creutz,
{\it Quarks, Gluons and Lattices},
(Cambridge Univ. Press, Cambridge, 1983).
%
\bibitem{Kogut}
J. Kogut, D.K. Sinclair, L. Susskind, Nucl. Phys. B114 (1976) 199.
%
\bibitem{Muenster}
G. M\"unster, Nucl. Phys. B190[FS3] (1981) 439 (E: B200[FS4] (1982) 536 and E: B205 (1982) 648).
%
\bibitem{Morningstar}
C. J. Morningstar and M.J. Peardon, Phys. Rev. D 60 (1999) 034509 [hep-lat/9901004]. 
%
\bibitem{Chen}
Y. Chen {\it et\ al.}, Phys. Rev. D 73 (2006) 014516 [hep-lat/0510074]. 
%
\bibitem{pavel2012}
H.-P. Pavel,
{\it Unconstrained Hamiltonian formulation of low-energy SU(3) Yang-Mills quantum theory},
arXiv: 1205.2237v1 [hep-th] (2012).
%
\bibitem{pavel2013}
H.-P. Pavel, 
PoS Confinement X (2012) 071, arXiv: 1303.3763 v1 [hep-th] (2013).
%
\bibitem{pavel2014}
H.-P. Pavel, 
 EPJ Web of Conferences {\bf 71} (2014) 00104 , 
arXiv: 1405.1970v1 [hep-th] (2014).
%
\bibitem{pavel2021}
H.-P. Pavel,
{\it Low-energy spectrum of SU(3) Yang-Mills Quantum Mechanics},
arXiv: 2112.06248 [hep-th] (2021).
%
\bibitem{Green and Gutperle}
M.B. Green and M. Gutperle, 
%
 Phys. Lett. {\bf B\ 398} (1997) 69.
\bibitem{Salgado}
F. Canfora, F. de Micheli, P. Salgado-Reolledo, and J. Zanelli,
Phys. Rev. D90 (2014) 044065.
%
\bibitem{Drell}
S.D. Drell, M. Weinstein and S. Yankielowicz,
Phys. Rev. D 14 (1976) 487, 1627.
%
\bibitem{J.Smit}
L.H. Karsten and J .Smit,
Phys.Lett 85 B (1979) 100.
\bibitem{Dittner}
P. Dittner, Commun. Math. Phys. 27 (1972) 44.
%
\bibitem{Simon}
B. Simon, Ann. Phys. 146 (1983) 209.
%
\bibitem{Nedelko and Voronin}
S. N. Nedelko and V.E. Voronin, 
Phys. Rev. D93 (2016) 094010.
%
%
\bibitem{Nair}
D. Karabali and V. P. Nair, Nucl. Phys. B 464 (1996) 135 [hep-th/9510157]; 
D. Karabali and V. P. Nair, Phys. Lett. B 379 (1996) 141 [hep-th/9602155]; 
D. Karabali, C. J. Kim and V. P. Nair, Nucl. Phys. B 524 (1998) 661 [hep-th/9705087].
\end{thebibliography}
\end{document}